\documentclass[10pt,journal,twoside]{IEEEtran}

\usepackage{cite}
\usepackage{theorem,bm,amssymb,amsmath,mathrsfs}
\usepackage{paralist}
\usepackage{tikz}
\newcommand*\circled[1]{\tikz[baseline=(char.base)]{
            \node[shape=circle,draw,inner sep=2pt] (char)
                 {$\scriptstyle #1$};}}
\def\graphedge{\mskip-1.5mu\mathchar"0200\mskip-3mu\mathchar"0200\mskip-1.5mu}
\def\graphleftedge{\mskip-1.25mu\mathchar"0220\mskip-3mu\mathchar"0200\mskip-1.5mu}
\def\graphrightedge{\mskip-1.5mu\mathchar"0200\mskip-3mu\mathchar"0221\mskip-1.25mu}

\usepackage{color}
\usepackage{graphicx}
\RequirePackage[pdftex,pdfpagemode=none, pdftoolbar=true,
                pdffitwindow=true,pdfcenterwindow=true]{hyperref}

\definecolor{darkgreen}{rgb}{0,0.5,0}
\definecolor{darkyellow}{rgb}{0.5,0.5,0}
\definecolor{darkred}{rgb}{0.6667,0,0}
\definecolor{light-gray}{gray}{0.5}

\makeatletter

\let\IfDraftVersion\if@draftclsmode

\def\threeauthors#1#2#3#4#5#6{\gdef\@address{}
   \gdef\@name{\begin{tabular}{@{}c@{}}
        {\em #1}\\\noalign{\vskip 6pt plus 3pt minus 3pt}
        #2\relax
   \end{tabular}\hskip .5in plus.5in minus.125in\begin{tabular}{@{}c@{}}
        {\em #3}\\\noalign{\vskip 9pt plus 3pt minus 3pt}
        #4\relax
   \end{tabular}\hskip .5in plus.5in minus.125in\begin{tabular}{@{}c@{}}
        {\em #5}\\\noalign{\vskip 6pt plus 3pt minus 3pt}
        #6\relax
\end{tabular}}}

\def\Rbbb{{\mathbb{R}}}

\def\Psib{{\bm{\Psi}}}
\def\Pib{{\bm{\Pi}}}

\def\Thetab{{\bm{\Theta}}}

\def\gammab{{\bm{\gamma}}}

\def\phib{{\bm{\phi}}}
\def\pib{{\bm{\pi}}}
\def\psib{{\bm{\psi}}}

\def\xib{{\bm{\xi}}}

\let\threat\theta
\let\threatb\thetab
\let\Threat\Theta

\def\Ab{{\mathbf{A}}}
\def\Bb{{\mathbf{B}}}

\def\Db{{\mathbf{D}}}
\def\Eb{{\mathbf{E}}}
\def\Hb{{\mathbf{H}}}
\def\Gb{{\mathbf{G}}}
\def\Ib{{\mathbf{I}}}

\def\Kb{{\mathbf{K}}}
\def\Lb{{\mathbf{L}}}
\def\Mb{{\mathbf{M}}}

\def\Qb{{\mathbf{Q}}}
\def\Rb{{\mathbf{R}}}
\def\Sb{{\mathbf{S}}}
\def\Tb{{\mathbf{T}}}
\def\Ub{{\mathbf{U}}}

\def\Wb{{\mathbf{W}}}
\def\Xb{{\mathbf{X}}}

\def\db{{\mathbf{d}}}

\def\lb{{\mathbf{l}}}

\def\sb{{\mathbf{s}}}

\def\zb{{\mathbf{z}}}

\def\pbmit{{\bm{p}}}

\def\zerob{{\bm{0}}}
\def\oneb{{\bm{1}}}

\def\LPolishb{\mbox{\bf\L}}

\DeclareMathOperator{\rank}{rank}

\DeclareMathOperator{\diag}{diag}
\DeclareMathOperator{\Diag}{Diag}

\DeclareMathOperator{\PD}{PD}
\DeclareMathOperator{\PFA}{PFA}
\DeclareMathOperator{\degree}{degree}
\DeclareMathOperator{\vertexsp}{\mathscr{V}}
\DeclareMathOperator{\Avg}{Avg}
\DeclareMathOperator{\dist}{dist}

\DeclareMathOperator{\avgpathlength}{\mit l}

\mathchardef\colonord="003A

\def\T{{\scriptscriptstyle\rm T}}   
\let\humlaut=\H
\def\H{\ifmmode{\scriptscriptstyle\rm H}\else\humlaut\fi}   

\def\by{\ifmmode $\hbox{-by-}$\else \leavevmode\hbox{-by-}\fi}
\def\sqrtm1{{\sqrt{\!-1}}}
\let\(=\langle
\let\)=\rangle

\def\detthresh{\mathrel{\mathop{\gtrless}\limits_{{\rm H}_0}^{{\rm H}_1}}}
\def\detthreshv#1{\mathrel{\mathop{\gtrless}\limits_{{\rm H}_0(#1)}^{{\rm H}_1(#1)}}}

\def\hgfarg#1{\left(\null\vcenter{\normalbaselines\m@th
    \ialign{&$\displaystyle##$\hfil\crcr
      \mathstrut\crcr\noalign{\kern-\baselineskip}
      #1\crcr\mathstrut\crcr\noalign{\kern-\baselineskip}}}\right)}
\def\Heqalign#1{\null\,\vcenter{\openup\jot\m@th
  \ialign{\strut\hfil$##$:\quad&\hfil$\displaystyle{##}$&$\displaystyle
      {{}##}$\hfil&\qquad##\hfil\crcr#1\crcr}}\,}
\def\eqalignno#1{\displ@y \tabskip\@centering
  \halign to\displaywidth{\hfil$\@lign\displaystyle{##}$\tabskip\z@skip
    &$\@lign\displaystyle{{}##}$\hfil\tabskip\@centering
    &\llap{$\@lign##$}\tabskip\z@skip\crcr
    #1\crcr}}
\def\dbleqalignno#1{\displ@y \tabskip\@centering
  \halign to\displaywidth{\hfil$\@lign\displaystyle{##}$\tabskip\z@skip
    &$\@lign\displaystyle{{}##}$\hfil
    &$\@lign\displaystyle{{}##}$\hfil\tabskip\@centering
    &\llap{$\@lign##$}\tabskip\z@skip\crcr
    #1\crcr}}

\theoremstyle{plain}
\newtheorem{definition}{\bf Definition}

\newtheorem{theorem}{\bf Theorem}

\newtheorem{problem}{\bf Problem}

\newif\ifmathtomb \mathtombfalse

\def\tombstone{\unskip\penalty50   
  \hskip 0pt plus-1fill \null\nobreak\hskip 0pt plus1fill
  \enskip \vrule width.3333em height.7em depth.2em
  \ifmmode \global\mathtombtrue \else \global\mathtombfalse \fi}

\newenvironment{hproof}
  {\futurelet\next\hpr@oftext}
  {\ifmathtomb \else \tombstone \fi \widowpenalty=10000  
   \par \ifmathtomb \else \addvspace{\medskipamount}\fi \global\mathtombfalse}
\def\hpr@oftext{\ifx\next[\let\temp\ohpr@@ftext\else\let\temp\hpr@@ftext\fi\temp}
\def\hpr@@ftext{\beginhpr@@f{Proof}}
\def\ohpr@@ftext[#1]{\beginhpr@@f{#1}}
\def\beginhpr@@f#1{\par \addvspace{\bigskipamount}%
  \noindent{\bf #1:\enspace}\ignorespaces }

\makeatother

\begin{document}


\let\originalnormalsize=\normalsize

\title{Bayesian Discovery of Threat Networks}

\author{\IEEEauthorblockN{Steven Thomas~Smith,~\IEEEmembership{Senior Member, IEEE}, Edward
  K. Kao,~\IEEEmembership{Member, IEEE},\\
  Kenneth~D. Senne,~\IEEEmembership{Life Fellow, IEEE}, Garrett Bernstein, and Scott Philips}
\thanks{Manuscript received November 15, 2013; revised March 17, 2014;
  accepted May 29, 2014. Date of publication July 08, 2014; date of
  current version September 8, 2004. The associate editor coordinating
  the review of this manuscript and approving it for publication was
  Prof.\ Francesco Verde. This work is sponsored by the Assistant
  Secretary of Defense for Research~\&\ Engineering under Air Force
  Contract FA8721-05-C-0002. Opinions, interpretations, conclusions
  and recommendations are those of the author and are not necessarily
  endorsed by the United States Government.}
\thanks{S.~T. Smith, K.~D. Senne, G.~Bernstein, and S.~Philips are
  with the MIT Lincoln Laboratory, Lexington, MA 02420 USA (e-mail:
  \href{mailto:stsmith@ll.mit.edu}{stsmith@ll.mit.edu};
  \href{mailto:edward.kao@ll.mit.edu}{edward.kao@ll.mit.edu};
  \href{mailto:senne@ll.mit.edu}{senne@ll.mit.edu};
  \href{mailto:garrett.bernstein@ll.mit.edu}{garrett.bernstein@ll.mit.edu}).}
\thanks{E.~K. Kao is with the MIT Lincoln Laboratory, Lexington, MA
  02420 USA, and also with the Department of Statistics, Harvard
  University; Cambridge MA USA 02138 (e-mail:
  \href{mailto:edwardkao@fas.harvard.edu}{edwardkao@fas.harvard.edu}).}
\thanks{Color versions of one or more of the figures in this paper are
  available online at \href{http://ieeexplore.ieee.org}{http://ieeexplore.ieee.org}.}

\thanks{Digital Object Identifier 10.1109/TSP.2014.2336613}}

\markboth{IEEE TRANSACTIONS ON SIGNAL PROCESSING, VOL. 62, NO. 20, 2014}%
{SMITH \MakeLowercase{\textit{et al.}}: BAYESIAN DISCOVERY OF THREAT NETWORKS}

\IEEEpubid{\lower3ex\vbox{\halign{\hss##\hss\cr 1053-587X~\copyright~2014 IEEE. Translations and content
  mining are permitted for academic research only. Personal use is
  also permitted, but republication/\cr redistribution requires IEEE
  permission. See
  \href{http://www.ieee.org/publications_standards/publications/rights/index.html}{http://www.ieee.org/publications\char`\_standards/publications/rights/index.html}
  for more information.\cr}}}

\setcounter{page}{5324}

\maketitle

\begin{abstract}A novel unified Bayesian framework for network
  detection is developed, under which a detection algorithm is derived
  based on random walks on graphs. The algorithm detects threat
  networks using partial observations of their activity, and is proved
  to be optimum in the Neyman-Pearson sense.  The algorithm is defined
  by a graph, at least one observation, and a diffusion model for
  threat. A link to well-known spectral detection methods is provided,
  and the equivalence of the random walk and harmonic solutions to the
  Bayesian formulation is proven.  A general diffusion model is
  introduced that utilizes spatio-temporal relationships between
  vertices, and is used for a specific space-time formulation that
  leads to significant performance improvements on coordinated covert
  networks.  This performance is demonstrated using a new hybrid
  mixed-membership blockmodel introduced to simulate random covert
  networks with realistic
  properties.
\end{abstract}

\IEEEkeywords Network detection, optimal detection, maximum likelihood
detection, community detection, network theory (graphs), graph theory,
diffusion on graphs, random walks on graphs, dynamic network models,
Bayesian methods, harmonic analysis, eigenvector centrality, Laplace
equations. \endIEEEkeywords

\overfullrule 5pt
\sloppy

\section{Introduction}
\label{sec:intro}  

\IEEEPARstart{N}{etwork} detection is the objective in many diverse
graph analytic applications, ranging from graph partitioning, mesh
segmentation, manifold learning, community
detection~\cite{Newman2006}, network anomaly
detection~\cite{Carter2010,Sandryhaila2014}, and the discovery of
clandestine
networks~\cite{Krebs2002,Neville2005,Sageman2004,Smith2011,Xu2008}.  A
new Bayesian approach to network detection is developed and analyzed
in this paper, with specific application to detecting small, covert
networks embedded within much larger background networks.  The novel
approach is based on a Bayesian probabilistic framework where the
probability of threat is derived from an observation model and an
a~priori threat diffusion model. Specifically, observed threats from
one or more vertices are propagated through the graph using a model
based on random walks represented as Markov chains with absorbing
states.  The resulting network detection algorithm is proved to be
optimum in the Neyman--Pearson sense of maximizing the probability of
detection at a fixed false alarm probability.  In the specific case of
space-time graphs with time-stamped edges, a model for threat
diffusion yields the new space-time threat propagation algorithm,
which is shown to be an optimal detector for covert networks with
coordinated activity.

Network detectors are analyzed using both a stochastic framework of
random walks on the graph and a probabilistic framework.  The two
frameworks are shown to be equivalent, providing an original, unified
approach for Bayesian network detection.  Performance for a variety of
Bayesian network detection algorithms is shown with both a stochastic
blockmodel and a new hybrid mixed-membership blockmodel (HMMB)
introduced to simulate random covert networks with realistic
properties.

\IEEEpubidadjcol

Using insights from algebraic graph theory, the connection between
this unified framework and other spectral-based network detection
methods~\cite{Donath1973,Fiedler1975,Newman2006} is shown, and the two
approaches are contrasted by comparing their different optimality
criteria based on detection probability and subgraph connectivity
properties. The random walk framework provides a connection with many
other well-known graph analytic methods that may also be posed in this
context~\cite{Avrachenkov2013,Chakrabarti2008,ChungZhao2010,Leskovec06,Onnela12,Shah2009,VanMieghem2011}. In
contrast to other research on network detection, rather than using a
sensor network to detect
signals~\cite{Chamberland2003,Alanyali2004,Sandryhaila2014}, the signal of
interest in this paper {\em is\/} the network. In this sense the paper
is also related to work on so-called manifold learning
methods~\cite{Belkin2003,Costa2004,Carter2010}, although the network
to be detected is a subgraph of an existing network, and therefore the
methods described here belong to a class of network anomaly
detection~\cite{Carter2010} as well as maximum-likelihood methods for
network detection~\cite{Ferry2009}.

Threat network discovery is predicated on the existence of
observations of network relationships.  Detection of network
communities is most likely to be effective if the communities exhibit
high levels of connection activity. The covert networks of interest in
this paper exist to accomplish nefarious, illegal, or terrorism goals,
while ``hiding in plain sight''~\cite{Xu2008}. Covert networks
necessarily adopt operational procedures to remain hidden and robustly
adapt to losses of parts of the
network~\cite{Carley2004,Sageman2004,Trinquier2006,Watts1999}.

This paper's major contributions are organized into a description of
the novel approach to Bayesian network detection in
Section~\ref{sec:Bayesnetdet}, and showing and comparing detection
performance using simulations of realistic networks in
Section~\ref{sec:modeling}. Fundamental new results are established in
Theorems
\ref{thm:maxprincthreatprop}--\ref{thm:prob-stochastic-equal}, which
prove a maximum principal for threat propagation, provide a
nonnegative basis for the principal invariant subspace, and prove the
equivalence between the probabilistic and stochastic realization
approaches of threat propagation. The Neyman--Pearson optimality of
threat propagation is established in Theorem~\ref{thm:tp-optimum}.

\section{Background}
\label{sec:background}

\subsection{Notation}
\label{sec:notation}

A graph~$G=(V,E)$ is defined by two sets, the vertices $V$, and the
edges $E\subset[V]^2$, in which $[V]^2$ denotes the set of $2$-element
subsets of~$V$~\cite{Diestel2000}. For example, the sets
$V=\{\,1,\;2,\;3\,\}$, $E=\bigl\{\,\{1,\,2\},\;\{2,\,3\}\,\bigr\}$
describe a simple graph with undirected edges between vertices $1$
and~$2$, and $2$ and~$3$:
$\circled{1}\graphedge\circled{2}\graphedge\circled{3}$. The {\em
  order\/} and {\em size\/} of~$G$ are defined to be $\#V$ and~$\#E$,
respectively.  A {\em subgraph} ${G'\subseteq G}$ is a graph $(V',E')$
with ${V'\subseteq V}$ and ${E'\subseteq E}$. If $E'$ contains all
edges in~$E$ with both endpoints in~$V'$, then $G'=G[V']$ is the {\em
  induced subgraph\/} of~$V'$. The {\em adjacency matrix\/}
${\Ab=\Ab(G)}$ of~$G$ is the $\{0,1\}$-matrix with ${a_{ij}=1}$ iff
$\{\,i,j\,\}\in E$. In the example, $\Ab= \left(\begin{smallmatrix}
  0&1&0\\ 1&0&1\\ 0&1&0\end{smallmatrix}\right)$. The adjacency matrix
  of simple or undirected graphs is necessarily symmetric.  The {\em
    degree matrix\/} ${\Db=\Diag(\Ab{\cdot}\oneb)}$ is the diagonal
  matrix of the vector of degrees of all vertices, where
  ${\oneb=(1,\ldots,1)^\T}$ is the vector of all ones. The {\em
    neighborhood\/} $N(u) =\bigl\{\,v:{\{u,v\}\in E}\,\bigr\}$ of a
  vertex~${u\in V}$ is the set of vertices adjacent to~$u$, or
  equivalently, the set of nonzero elements in the $u$-th row of~$\Ab$.
  The {\em vertex space\/} $\vertexsp(G)$ of~$G$ is the vector space
  of functions $f\colon V\to\{0,1\}$.

A {\em directed graph\/} $G_\sigma$ is defined by an orientation map
$\sigma\colon[V]^2\to {V\times V}$ (the ordered Cartesian product
of~$V$ with itself) in which the first and second coordinates are
called the initial and terminal vertices, respectively. A {\em
  strongly connected graph\/} is a connected graph for which a
directed path exists between any two vertices.  The {\em incidence
  matrix\/} ${\Bb=\Bb(G_\sigma)}$ of~$G_\sigma$ is the
$(0,\pm1)$-matrix of size $\#V\by\#E$ with ${\Bb_{ie}=\pm1}$, if $i$
is an terminal\slash initial vertex of~$\sigma(e)$, and $0$
otherwise. For example, the directed graph
$\circled{1}\graphleftedge\circled{2}\graphrightedge\circled{3}$ has
incidence matrix $\Bb= \left(\begin{smallmatrix}
  \hphantom{-}1&\hphantom{-}0\\ -1&-1\\ \hphantom{-}0&\hphantom{-}1\end{smallmatrix}\right)$.
  The {\em unnormalized Laplacian matrix\/} or {\em Kirchhoff
    matrix\/} $\Qb$ of a graph, the (normalized) {\em Laplacian
    matrix\/} $\Lb$, and the {\em generalized\/} or {\em asymmetric
    Laplacian matrix\/} $\LPolishb$ are, respectively,
\begin{align} \Qb&=\Bb\Bb^\T=\Db-\Ab,\label{eq:kirchhoff}\\ \Lb
    &=\Db^{-1/2}\Qb\Db^{-1/2}=\Ib-\Db^{-1/2}\Ab\Db^{-1/2},\label{eq:laplacian}\\ \LPolishb
  &=\Db^{-1/2}\Lb\Db^{1/2} =\Db^{-1}\Qb
  =\Ib-\Db^{-1}\Ab. \label{eq:genlaplacian}\end{align} In the example,
$\LPolishb$ is immediately recognized as a discretization of the
second derivative $-d^2\!/\!dx^2$, i.e.\ the negative of the
\hbox{$1$-d} Laplacian operator $\Delta =\partial^2\!/\partial x^2
+\partial^2\!/\partial y^2+\cdots$ that appears in physical
applications. The connection between the Laplacian matrices and
physical applications is made through Green's first identity, a link
that explains many theoretical and performance advantages of the
normalized Laplacian over the Kirchhoff matrix across
applications~\cite{Chung1994,Weiss1999,vonLuxburg2005,White2005}.

Solutions to Laplace's equation on a graph are directly connected to
random walks or discrete Markov chains on the vertices of the graph,
which provide stochastic realizations for harmonic problems. A {\em
  (right) stochastic matrix\/}~$\Tb$ of a graph is a nonnegative
matrix such~that ${\Tb\oneb=\oneb}$. This represents a state
transition matrix of a random walk on the graph with transition
probability $t_{ij}$ of jumping from vertex~$v_i$ to
vertex~$v_j$. The Perron--Frobenius theorem guarantees if $\Tb$ is
irreducible (i.e.\ $G$ is strongly connected) then there exists a
stationary probability distribution $\pbmit_v$ on~$V$ such~that
${\pbmit^\T\Tb=\pbmit^\T}$~\cite{Gantmacher59,Godsil2001}. Random walk
realizations can be used to describe the solution to harmonic boundary
value problems, e.g.\ equilibrium
thermodynamics~\cite{Ozisik1968,Sabelfeld1994}, in which given values
are proscribed at specific ``boundary'' vertices.

\subsection{Network Detection}
\label{sec:connections}

Network detection is a special class of the more general graph
partitioning (GP) problem in which the binary decision of membership
or non-membership for each graph vertex must be determined. Indeed,
the network detection problem for a graph $G$ of order~$N$ results in
a $2^N$-ary multiple hypothesis test over the vertex space
$\vertexsp(G)$, and, when detection optimality is considered, an
optimal test involves partitioning the measurement space into $2^N$
regions yielding a maximum probability of detection (PD).  This
NP-hard combinatoric problem is computationally and analytically
intractable. In general, network detection methods invoke various
relaxation approaches to avoid the NP-hard network detection problem.
The new Bayesian threat propagation approach taken in this paper is to
greatly simplify the general $2^N$-ary multiple hypothesis test by
applying the random walk model and treating it as $N$ independent
binary hypothesis tests. This approach is related to existing network
detection methods by posing an optimization problem on the
graph---e.g.\ threat propagation maximizes PD---and through solutions
to Laplace's equation on graphs.  Because many network detection
algorithms involve such solutions, a key fact is that the constant
vector ${\oneb=(1,\ldots,1)^\T}$ is in the kernel of the
Laplacian, \begin{equation}\Qb\oneb=\zerob;\qquad
  \LPolishb\oneb=\zerob.\label{eq:Lonekernel}\end{equation} This
constant solution does not distinguish between vertices at all, a
deficiency that may be resolved in a variety of ways.

Efficient graph partitioning algorithms and analysis appeared in the
1970s with Donath and Hoffman's eigenvalue-based bounds for graph
partitioning~\cite{Donath1973} and Fiedler's connectivity analysis and
graph partitioning algorithm~\cite{Fiedler1975} which established the
connection between a graph's algebraic properties and the spectrum of
its Kirchhoff Laplacian matrix~${\Qb=\Db-\Ab}$
[Eq.~(\ref{eq:kirchhoff})]. Spectral methods solve the graph
partitioning problem by optimizing various subgraph connectivity
properties. Similarly, the threat propagation algorithm developed here
in Section~\ref{sec:Bayesnetdet} optimizes the probability of
detecting a subgraph for a specific Bayesian model. Though the
optimality criteria for spectral methods and threat propagation are
different, all these network detection methods must address the
fundamental problem of avoiding the trivial solution of constant
harmonic functions on graphs.  Threat propagation avoids this problem
by using observation vertices and a~priori probability of threat
diffusion (Section~\ref{sec:sptp}). Spectral methods take a
complementary approach to avoid this problem by using an alternate
optimization criterion that depends upon the network's topology.

The {\em cut size\/} of a subgraph---the number of edges necessary to
remove to separate the subgraph from the graph---is quantified by the
quadratic form~$\sb^\T\Qb\sb$, where $\sb=(\pm1,\ldots,\pm1)^\T$ is a
$\pm1$-vector who entries are determined by subgraph
membership~\cite{Pothen1990}. Minimizing this quadratic form
over~$\sb$, whose solution is an eigenvalue problem for the graph
Laplacian, provides a network detection algorithm based on the model
of minimal cut size. However, there is a paradox in the application of
spectral methods to network detection: the smallest eigenvalue of the
graph Laplacian ${\lambda_0(\Qb)=0}$ corresponds to the
eigenvector~$\oneb$ constant over~all vertices, which fails to
discriminate between subgraphs. Intuitively this degenerate constant
solution makes sense because the two subgraphs with minimal (zero)
subgraph cut size are the entire graph itself (${\sb\equiv\oneb}$), or
the null graph (${\sb\equiv-\oneb}$). This property manifests itself
in many well-known results from complex analysis, such as the maximum
principle.

Fiedler showed that if rather the eigenvector~$\xib_1$ corresponding
to the second smallest eigenvalue $\lambda_1(\Qb)$ of~$\Qb$ is used
(many authors write ${\lambda_1=0}$ and~$\lambda_2$ rather than the
zero offset indexing ${\lambda_0=0}$ and~$\lambda_1$ used here), then
for every nonpositive constant~${c\le 0}$, the subgraph whose vertices
are defined by the threshold ${\xib_1\ge c}$ is necessarily
connected. This algorithm is called {\em spectral detection}.  Given a
graph~$G$, the number $\lambda_1(\Qb)$ is called the {\em Fiedler
  value\/} of~$G$, and the corresponding eigenvector $\xib_1(\Qb)$ is
called the {\em Fiedler vector}. Completely analogous with comparison
theorems in Riemannian geometry that relate topological properties of
manifolds to algebraic properties of the Laplacian, many graph
topological properties are tied to its Laplacian. For example, the
graph's diameter~$D$ and the minimum degree $d_{\rm min}$ provide
lower and upper bounds for the Fiedler value $\lambda_1(\Qb)$:
$4/(nD)\le \lambda_1(\Qb)\le n/({n-1}){\cdot}d_{\rm
  min}$~\cite{Mohar1991}. This inequality explains why the Fiedler
value is also called the {\em algebraic connectivity\/}: the greater the
Fiedler value, the smaller the graph diameter, implying greater graph
connectivity. If the normalized Laplacian $\Lb$ of
Eq.~(\ref{eq:laplacian}) is used, the corresponding inequality
involving the generalized eigenvalue
${\lambda_1(\Lb)=\lambda_1(\Qb,\Db)}$ involves the graph's
diameter~$D$ and volume~$V$: $1/(DV)\le \lambda_1(\Lb)\le
n/({n-1})$~\cite{Chung1994}.

Because in practice spectral detection with its implicit assumption of
minimizing the cut size oftentimes does not detect intuitively
appealing subgraphs, Newman introduced the alternate criterion of
subgraph ``modularity'' for subgraph detection~\cite{Newman2006}.
Rather than minimize the cut size, Newman proposes to maximize the
subgraph connectivity relative to background graph connectivity, which
yields the quadratic maximization problem~$\max_\sb\sb^\T\Mb\sb$,
where ${\Mb=\Ab-V^{-1}\db\db^\T}$ is Newman's {\em modularity matrix},
$\Ab$ is the adjacency matrix, ${(\db)_i=d_i}$ is the degree vector,
and $V=\oneb^\T\db$ is the graph volume~\cite{Newman2006}. Newman's
modularity-based graph partitioning algorithm, also called community
detection, involves thresholding the values of the principal
eigenvector of~$\Mb$. Miller
et~al.~\cite{Miller2011,Miller2010a,Miller2010b} also consider
thresholding arbitrary eigenvectors of the modularity matrix, which by
the Courant minimax principle biases the Newman community detection
algorithm to smaller subgraphs, a desirable property for many
applications. They also outline an approach for exploiting
observations within the spectral framework~\cite{Miller2011}.

Other graph partitioning methods invoke alternate relaxation
approaches that yield practical detection\slash partitioning
algorithms such as semidefinite programming
(SDP)~\cite{Wolkowicz1999,Arora2008,Leskovec2010}.  A class of graph
partitioning algorithms is based on infinite random walks on
graphs~\cite{Spielman2004}. Zhou and Lipowsky define proximity using
the average distance between vertices~\cite{Zhou2004}.  Anderson
et~al.\ define a local version biased towards specific
vertices~\cite{Andersen2006}. Mahoney et~al.\ develop a local spectral
partitioning method by augmenting the quadratic optimization problem
with a locality constraint and relaxing to a convex
SDP\null~\cite{Mahoney2012}.  An important dual to network detection
is the problem of identifying the source of an epidemic or rumor using
observations on the graph~\cite{Shah2011,Shah2009}. Another related
problem is the determination of graph topologies for which epidemic
spreading occurs~\cite{Chakrabarti2008,VanMieghem2011}.  The approach
adopted in this paper has fundamentally different objectives and
propagation models than the closely-related epidemiological problems.
These problems focus on disease spreading to large portions of the
entire graph, which arises because disease may spread from any
infected neighbor---yielding a logical OR of neighborhood disease.
Network detection focuses on discovering a subgraph most likely
associated with a set of observed vertices, assuming random walk
propagation to the observations---yielding an arithmetic mean of
neighborhood threat.  All of these methods are related to spectral
partitioning through the graph Laplacian.

\section{Bayesian Network Detection}
\label{sec:Bayesnetdet}

The Bayesian model developed here depends upon threat observation and
propagation via random walks over both space and time, and the
underlying probabilistic models that govern inference from observation
to threat, then propagation of threat throughout the graph.  Bayes'
rule is used to develop a network detection approach for spatial-only,
space-time, and hybrid graphs. The framework assumes a given Markov
chain model for transition probabilities, and hence knowledge of the
graph, and a diffusion model for threat.  Neyman--Pearson optimality is
developed in the context of network detection with a simple binary
hypothesis, and it is proved that threat propagation is optimum in
this sense.

The framework is sufficiently general to capture graphs formed by many
possible relationships between entities, from simple graphs with
vertices that represent a single type of entity, to bipartite or
multipartite graphs with heterogeneous entities. For example, an email
network is a bipartite graph comprised of two types of vertices:
individual people and individual email messages, with edges
representing a connection between people and messages.  Without loss
of generality, all entity types to be detected are represented as
vertices in the graph, and their connections are represented by edges
weighted by scalar transition probabilities.

Network detection is the problem of identifying a specific subgraph
within a given graph~${G=(V,E)}$. Assume that within $G$, a foreground
or ``threat'' network $V^\Theta$ exists defined by an (unknown) binary
random variable:

\begin{definition} {\em Threat\/} is a\/ $\{0,1\}$-valued
  discrete random variable. {\em Threat on a graph\/} ${G=(V,E)}$ is a
  $\{0,1\}$-valued function\/ ${\Theta\in\vertexsp(G)}$. Threat at the
  vertex~$v$ is denoted\/ $\Theta_v$. A vertex ${v\in V}$ is in the\/
  {\em foreground\/} if\/ ${\Theta_v=1}$, otherwise $v$ is in the\/ {\em
    background}. \label{def:threat}\end{definition}

The foreground or threat vertices are the set $V^\Theta
=\{\,\,v:{\Theta_v=1}\}$, and the foreground or threat network is the
induced subgraph $G^\Theta=G[V^\Theta]$. A network detector of the
subgraph $G^\Theta$ is a collection of binary hypothesis tests to
decide which of the graph's vertices belong to the foreground
vertices~$V^\Theta$. Formally, a network detector is an element of the
vertex space of~$G$:

\begin{definition} Let ${G=(V,E)}$ be a graph. A\/ {\em network
    detector\/}~$\phi$ on~$G$ is a\/ $\{0,1\}$-valued function
  ${\phi\in\vertexsp(G)}$. The induced subgraph $G^\phi=G[V^\phi]$ of
  $V^\phi=\{\,v:{\phi_v=1}\,\}$ is called the\/ {\em foreground network\/}
  and the induced subgraph $G^{\tilde\phi}=G[V^{\tilde\phi}]$ of
  $V^{\tilde\phi}=\{\,v:\phi_v=0\,\}$ is called the\/ {\em background
    network}, in which $\tilde\phi$ denotes the logical complement
  of~$\phi$.
\end{definition}

The correlation between a network detector $\phi$ and the actual
threat network defined by the function $\Theta$ determines the
detection performance of~$\phi$, measured using the detector's
probability of detection (PD) and probability of false alarm
(PFA). The PD and PFA of~$\phi$ are the fraction of correct and
incorrect foreground vertices determined by~$\phi$: \begin{align}
  \PD^\phi &=\#(V^\phi\cap V^\Theta)/\#V^\Theta,\\ \PFA^\phi
  &=\#(V^\phi\cap V^{\tilde\Theta})/\#V^{\tilde\Theta}.\end{align}

Observation models are now introduced and applied in the sequel to
threat propagation models in the contexts of spatial-only graphs,
space-time graphs whose edges have time stamps, and finally a hybrid
graphs with edges of mixed type. Assume that there are $C$ observed
vertices $\{\,v_{b_1},\ldots,v_{b_C}\,\}\subset V$ at which
observations are taken. In the resulting Laplacian problem, these are
``boundary'' vertices, and the rest are ``interior.''  The simplest
case involves scalar measurements; however, there is a straightforward
extension to multidimensional observations.

\begin{definition} Let ${G=(V,E)}$ be a graph. An\/ {\em observation
  on the graph\/} is a vector\/ $\zb\colon
  \{\,v_{b_1},\ldots,v_{b_C}\,\}\to M\subset\Rbbb^C$ from
  $C$~vertices to a\/ {\em measurement space\/}~$M\subset\Rbbb^C$.
\end{definition}

Ideally, observation of a foreground and\slash or background vertices
unequivocally determines whether the observed vertices lie in the
foreground or background networks, i.e.\ given a foreground graph
${G^\Theta=G[V^\Theta]}$ and a foreground vertex ${v\in V^\Theta}$, an
observation vector $\zb$ evaluated at~$v$ would yield $\zb(v)=1$, and
$\zb$ evaluated at a background vertex ${v'\in V^{\tilde\Theta}}$ would
yield $\zb(v')=0$. In general, it is assumed that the observation
$\zb(v)$ at~$v$ and the threat $\Theta_v$ at~$v$ are not statistically
independent, i.e.\ $f\bigl(\zb(v)\mid\Theta_v\bigr)\ne
f\bigl(\zb(v)\bigr)$ for probability density $f$, so that there is
positive mutual information between $\zb(v)$ and~$\Theta_v$. Bayes'
rule for determining how likely a vertex is to be a foreground member
or not depends on the model linking observations to threat:

\begin{definition} Let ${G^\Theta=G[V^\Theta]}$ be the foreground graph
of a graph~$G$ determined by\/ ${\Theta\in\vertexsp(G)}$, and let\/
$\zb\colon \{\,v_{b_1},\ldots,v_{b_C}\,\}\to M\subset\Rbbb^C$ be an
observation on~$G$. The conditional probability density
$f\bigl(\zb(v)\mid\Theta_v\bigr)$ is called the\/ {\em observation model}
of vertex ${v\in V}$.\label{def:om}\end{definition}

The simplest, ideal observation model equates threat with observation
so that $f_{\rm
  ideal}\bigl(\zb(v)\mid\Theta_v\bigr)=\delta_{\zb(v)\Theta_v}$ in which
$\delta_{ij}$ is the Kronecker delta. Though the threat network
hypotheses are being treated here independently at each vertex, this
framework allows for more sophisticated global models that include
hypotheses over two or more vertices.

The remainder of this section is devoted to the development of
Bayesian methods of using measurements on a graph to determine the
probability of threat on a graph in various contexts---spatial-only,
space-timed, and the hybrid case---then showing that these methods are
optimum in the Neyman--Pearson sense of maximizing the probability of
detection at a given false alarm rate. The motivating problem is:

\begin{problem} Detect the foreground graph ${G^\Theta=G[V^\Theta]}$ in
the graph ${G=(V,E)}$ with an unknown foreground
${\Theta\in\vertexsp(G)}$ and known observation vector
$\zb(v_{b_1},\ldots,v_{b_C})$.\label{prob:netdet}\end{problem}

This problem is addressed by computing the probability of threat
$P(\Theta_v)$ at all graph vertices from the measurements at observed
vertices using an observation model and the application of Bayes'
rule.

\subsection{Spatial Threat Propagation}
\label{sec:sptp}

A spatial threat propagation algorithm is motivated and developed now,
which will be used in the subsequent space-time generalization, and
will demonstrate the connection to spectral network detection
methods. A vertex is declared to be threatening if the observed threat
propagates to that vertex.  We wish to compute the probability of
threat $P(\Theta_v=1\mid\zb)$ at all vertices ${v\in V}$ in a graph
${G=(V,E)}$ given an observation $\zb(v_{b_1},\ldots,v_{b_C})$ on~$G$.
Implicit in Problem~\ref{prob:netdet} is a coordinated threat network
in which threat propagates via network connections, i.e.\ graph edges.
For simplicity, probabilities conditioned on the observation~$\zb$
will be written \begin{equation}\threat_v =P(\Theta_v\mid\zb)
  \label{eq:varthetav}\end{equation} with an implied
dependence on the observation vector~$\zb$ and the event
${\Theta_v=1}$ expressed as $\Theta_v$.

To model the diffusion of threat throughout the graph, we introduce an
a~priori probability~$\psi_v$ at each vertex~$v$ that represents
threat diffusion at~$v$. $\psi_v$ is the probability that threat
propagates through vertex~$v$ to its neighbors, otherwise threat
propagates to an absorbing ``non-threat'' state with probability
${1-\psi_v}$. A threat diffusion event at~$v$ is represented by the
${\{0,1\}}$-valued r.v.\ $\Psi_v$:

\begin{definition}\label{def:probtp} The\/ {\em threat
  diffusion model\/} of a graph ${G=(V,E)}$ with observation\/~$\zb$
  is given by the a~priori\/ ${\{0,1\}}$-valued event\/ $\Psi_v$ that
  threat\/~$\Theta_v$ propagates through~$v$ with probability
  $\psi_v$.\end{definition}

Threat propagation on the graph from the observed vertices to all
other vertices is defined as an average over all random walks between
vertices and the observations. A single random walk between~$v$ and an
observed vertex~$v_{b_c}$ is defined by the sequence
\begin{equation}\hbox{\rm walk}_{v\to
  v_{b_c}}=(v_{w_1},v_{w_2},\ldots,v_{w_L})\label{eq:walk}\end{equation}
with endpoints ${v_{w_1}=v}$ and ${v_{w_L}=v_{b_c}}$, comprised of $L$
steps along vertices ${v_{w_l}\in V}$.  The probabilities for each
step of the random walk are defined by the elements of the transition
matrix $t_{vu}$
from vertex~$v$ to~$u$, multiplied by the a~priori probability
$\psi_v$ that threat propagates through~$v$. The assumption that~$G$
is strongly connected guarantees the existence of a walk between every
vertex and every observation.  Threat may be absorbed to the
non-threat state with probability ${1-\psi_{v_{w_l}}}$ at each step.
The simplest models for both the transition and a~priori probabilities
are uniform: ${t_{ij}=1/\degree(v_i)}$ for ${(i,j)\in E}$,
i.e.\ ${\Tb=\Db^{-1}\Ab}$, and ${\psi_v\equiv1}$. The implications of
these simple models as well as more general weighted models will be
explored throughout this section.

The indicator function \begin{equation}I_{{\rm walk}_{v\to v_{b_c}}}
  =\prod\nolimits_{l=1}^L\Psi^{(l)}_{v_{w_l}}\label{eq:Iwalk}\end{equation}
determines whether threat propagates along the walk or is absorbed
into the non-threat state (the superscript `$(l)$' allows for the
possibility of repeated vertices in the sequence).  The definition of
threat propagation is captured in three parts: (1)~a single random
walk, $\hbox{\rm walk}_{v\to v_{b_c}}$, with ${I_{{\rm walk}_{v\to
      v_{b_c}}}=1}$ yields threat probability $\theta_{v_{b_c}}$
at~$v$; (2)~the probability of threat averaged over all such random
walks; (3)~the random variable obtained by averaging the
r.v.\ $\Theta_{v_{b_c}}$ over all such random walks. Formally,

\begin{figure}[t]
\medskip
\normalsize
\centerline{\includegraphics{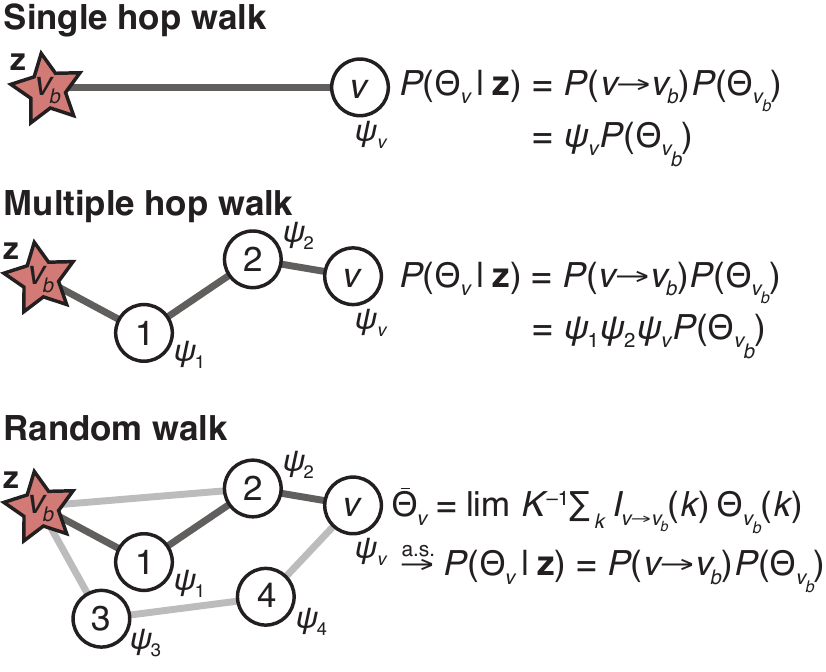}}
\caption{Illustration of the random walk representation for threat
  propagation from Definition~\ref{def:threatprop} and
  Eqs.\ (\ref{eq:Threat-v-walk}) and (\ref{eq:threat-walk}), for the
  case of a single observation. The upper illustration shows the
  simplest, trivial case with a single hop from the observation to the
  vertex. The middle illustration shows the next simplest case with
  multiple hops. The lower illustration shows an example of the
  general case, comprised of the simpler multiple hop
  case.\label{fig:randomwalk}}
\end{figure}

\begin{definition} {\bf(Threat Propagation).}\quad Let ${G=(V,E)}$ be a strongly
  connected graph with threat probabilities
  $\threat_{v_{b_1}}$,~\dots, $\threat_{v_{b_C}}$ at observed vertices
  $v_{b_1}$,~\dots, $v_{b_C}$ and the threat diffusion model $\psi_v$
  for~all ${v\in V}$. (1)~For a random walk on~$G$ from~$v$ to
  observed vertex $v_{b_c}$ with transition matrix\/~$\Tb$, $\hbox{\rm
    walk}_{v\to v_{b_c}}=(v_{w_1},v_{w_2},\ldots,v_{w_L})$, if
  events\/ ${\Psi_{v_{w_l}}\equiv1}$ for~all vertices~$v_{w_l}$ along
  the walk, then the\/ {\em threat propagation from~$v_{b_c}$ to~$v$
    along\/} $\hbox{\rm walk}_{v\to v_{b_c}}$ is defined to
  be~$\threat_{v_{b_c}}$; otherwise, the threat equals
  zero. (2)\/~{\em Threat propagation\/} to vertex~$v$ is defined as
  the expectation of threat propagation to~$v$ along all random walks
  emanating from~$v$, \begin{equation} \theta_v
    =\lim_{K\to\infty}{1\over K}\sum_k I_{{\rm walk}^{(k)}_{v\to
        v_{b_{c(k)}}}}\threat_{v_{b_{c(k)}}}, \label{eq:threat-v-walk}\end{equation}
  where the $k$th walk terminates at the observed vertex
  $v_{b_{c(k)}}$.  (3)\/~{\em Random threat propagation\/} to
  vertex~$v$ is defined as the random variable \begin{equation}
    \bar\Theta_v =\lim_{K\to\infty}{1\over K}\sum_k I_{{\rm
        walk}^{(k)}_{v\to
        v_{b_{c(k)}}}}\Threat^{(k)}_{v_{b_{c(k)}}} \label{eq:Threat-v-walk}\end{equation}
  with independent draws $\Threat^{(k)}_{v_{b_{c(k)}}}$ of the
  observed threat.\label{def:threatprop}\end{definition}

Fig.~\ref{fig:randomwalk} illustrates threat propagation of
Definition~\ref{def:threatprop} and Eq.~(\ref{eq:Threat-v-walk}) [and
  Eq.~(\ref{eq:threat-walk}) from the sequel] for the
simple-to-general cases of a single hop, multiple hops, and an
arbitrary random walk.  By the law of large
numbers, \begin{equation}{\bar\Threat_v\buildrel{\rm
      a.s.}\over\to\threat_v}\quad\hbox{as
    $K\to\infty$}.\label{eq:barTheta}
\end{equation}

The random walk model is described using the distinct yet equivalent
probabilistic and stochastic realization representations.  The
probabilistic representation describes the threat probabilities by a
Laplacian system of linear equations, which amounts to equating threat
at a vertex to an average of neighboring probabilities.  In contrast,
the stochastic realization representation presented below in
Section~\ref{sec:sptp-stochastic} describes the evolution of a single
random walk realization whose ensemble statistics are described by the
probabilistic representation, presented next.

\subsubsection{Probabilistic Approach}
\label{sec:sptp-prob}

Consider the (unobserved) vertex
$v\not\in\{\,v_{b_1},\ldots,v_{b_C}\,\}$ with neighbors
$N(v)=\{\,v_{n_1},\ldots,v_{n_{d_v}}\,\}\subset V$ and
$d_v=\degree(v)$. The probabilistic equation for threat propagation
from the neighbors of a vertex~$v$ follows immediately from
Definition~\ref{def:threatprop} from first-step analysis, yielding the
{\em threat propagation equation:\/} \begin{equation}\threat_v
  =\psi_v\sum\nolimits_{u\in
    N(v)}t_{vu}\threat_u,\label{eq:tp1}\end{equation} which is simply
the average of the neighboring threat probabilities weighted by
transition probabilities ${t_{vu}=(\Tb)_{vu}}$. Note that because
${\Ab\threatb\ge\threatb}$, $\threat_v$ is a subharmonic function on
the graph~\cite{Dynkin1969,Godsil2001}. In the simplest case of
uniform transition probabilities, ${\Tb=\Db^{-1}\Ab}$
and \begin{equation}\threat_v ={\psi_v\over d_v}\sum\nolimits_{u\in
    N(v)}\threat_u.\label{eq:tp2}\end{equation} Expressed in
matrix-vector notation, Eqs.\ (\ref{eq:tp1}) and~(\ref{eq:tp2})
become \begin{equation} \threatb =\Psib\Tb\threatb
  \quad\hbox{and}\quad \threatb
  =\Psib\Db^{-1}\Ab\threatb, \label{eq:tp3}\end{equation} where
${(\threatb)_v=\threat_v}$, ${\Psib=\Diag(\psi_v)}$ is the diagonal
matrix of a~priori threat diffusion probabilities, $\Tb$, $\Db$, and
$\Ab$ are, respectively, the transition, degree, and adjacency
matrices. The threat probabilities at the observed vertices
$v_{b_1}$,~\dots, $v_{b_C}$ are determined by the observation model of
Definition~\ref{def:om}, and threat probabilities at all other
vertices are determined by solving Eq.~(\ref{eq:tp3}), as with all
Laplacian boundary value problems.

As seen in the spectral network detection methods in
Section~\ref{sec:connections}, many network detection algorithms
exploit properties of the graph Laplacian, and therefore must address
the fundamental challenge posed by the implication of the maximum
principle that harmonic functions are constant~\cite{Dynkin1969} in
many important situations [Eq.~(\ref{eq:Lonekernel})], and because the
constant function does not distinguish between vertices, detection
algorithms that rely only on solutions to Laplace's equation provide a
futile approach to detection.  If the boundary is constant, i.e.\ the
probability of threat on all observed vertices is equal, then this is
the probability of threat on every vertex in the graph. The later
example is relevant in the practical case in which there a single
observation.  The maximum principle applies directly to threat
propagation with uniform prior ${\Psib=\Ib}$ and uniform probability
of threat~$p_o$ on the observed vertices: Eqs.~(\ref{eq:tp3}) are
recognized as Laplace's equation, ${(\Ib-\Tb)\threatb=\zerob}$ or
${(\Ib-\Db^{-1}\Ab)\threatb=\zerob}$, whose solution is trivially
${\threatb=p_o\oneb}$. Equivalently, from the stochastic realization
point-of-view, the probability of threat on all vertices is the same
because average over all random walks between any vertex to a boundary
(observed) vertex is trivially the observed, constant probability of
threat~$p_o$.

The following maximum principle establishes the existence of a unique
non-negative threat probability on a graph given threat probabilities
at observed vertices:

\begin{theorem} {\bf(Maximum Principle for Threat Propagation).}\quad Let
  $G=(V,E)$ be a connected graph with positive probability of threat
  $\threat_{v_{b_1}}$,~\dots, $\threat_{v_{b_C}}$ at observed vertices
  $v_{b_1}$,~\dots, $v_{b_C}$ and the a~priori probability $\psi_v$
  that threat propagates through vertex~$v$. Then there exists a
  unique probability of threat $\threat_v$ at all vertices such that
  ${\threat_v\ge0}$ and the maximum threat occurs at the observed
  vertices. \label{thm:maxprincthreatprop}
\end{theorem}

\begin{hproof} That $\threat_v$ exists follows from the connectivity
  of~$G$, and that it takes its maximum on the boundary follows
  immediately from Eq.~(\ref{eq:tp2}) because the threat at all
  vertices is necessarily bounded above by their neighbors. Now prove
  that $\threat_v$ is nonnegative by establishing a contradiction. Let
  $\threat_m$ be the minimum of~all ${\threat_v<0}$. Because
  $\psi_m\le1$, Eq.~(\ref{eq:tp2}) implies that $\threat_m\ge
  \Avg[N(m)]$, the weighted average value of the neighbors
  of~$m$. Therefore, there exists a neighbor ${n\in N(m)}$ such~that
  ${\threat_n\le\threat_m}$. But $\threat_m$ is by assumption the
  minimum value. Therefore, ${\threat_n=\threat_m}$ for~all ${n\in
    N(m)}$. Because $G$ is connected, ${\threat_v\equiv\threat_m}$ is
  constant for all unobserved vertices on~$G$. Now consider the
  minimum threat $\threat_i$ for which ${i\in N(b)}$ is a neighbor
  of an observed vertex~$b$. By
  Eq.~(\ref{eq:tp2}), \begin{align}\threat_i
    &=\textstyle\psi_id_i^{-1}\Bigl(\sum_{j\in N(i)\backslash
      b}\threat_j +\threat_{b}\Bigr),\\ &\ge
    \psi_id_i^{-1}\bigl((d_i-1)\threat_i+\threat_{b}\bigr).
  \end{align} Therefore, \begin{equation} \threat_i\ge
    {\psi_i\threat_{b}\over (1-\psi_i)d_i+\psi_i}\ge 0,\end{equation}
  a contradiction. Therefore, the minimum value of~$\threat_v$ is
  nonnegative.
\end{hproof}

This theorem is intuitively appealing because it shows how nonuniform
a~priori probabilities~$\psi_v$ yield a nonconstant and nonnegative
threat on the graph; however, the theorem conceals the crucial
additional ``absorbing'' state that allows threat to dissipate away
from the constant solution. This slight defect will be corrected
shortly when the equivalent stochastic realization Markov chain model
is introduced. Models about the likelihood of threat at specific
vertices across the graph are provided by the a~priori probabilities
$\psi_v$, which as discussed above prevent the uninformative (yet
valid) solution of constant threat across the graph given an
observation of threat at a specific vertex. 

A simple model for the a~priori probabilities is degree-weighted
threat propagation (DWTP), \begin{equation}\psi_v={1\over d_v}\qquad
  \hbox{(DWTP)} ,\label{eq:dwtp}\end{equation} in which threat is less
likely to propagate through high-degree vertices. Another simple model
sets the mean propagation length proportional to the graph's average
path length~$\avgpathlength(G)$ yields length-weighted threat
propagation (LWTP) \begin{equation}\psi_v\equiv
  2^{-1/\avgpathlength(G)}\qquad\hbox{(LWTP)}. \label{eq:lwtp}\end{equation}
For almost-surely connected Erd\humlaut{o}s--R\'enyi graphs with
${p=n^{-1}\log n}$, $l(G) = ({\log n-\gamma})/\log\log n+1/2$ and
${\gamma=0.5772\ldots}$ is Euler's constant~\cite{Fronczak2004}. A
model akin to breadth-first search (BFS) sets the a~priori
probabilities to be inversely proportional to the Dijkstra distance
from observed vertices, i.e.\ \begin{equation}\psi_v\propto
1/\dist(v,\{\,v_{b_1},\ldots,v_{b_C}\,\})\qquad\hbox{(BFS)}. \label{eq:psi-bfs}\end{equation}

Defining the generalized Laplacian
operator \begin{equation}{\LPolishb^\psib \buildrel{\rm def}\over
    =\Ib-\Psib\Db^{-1}\Ab},\label{eq:Lpsib}\end{equation} the threat
propagation equation Eq.~(\ref{eq:tp3}) written
as \begin{equation}\LPolishb^\psib\threatb=\zerob, \label{eq:sptp-gl}\end{equation}
connects the generalized asymmetric Laplacian matrix of with threat
propagation, the solution of which itself may be viewed as a boundary
value problem with the harmonic operator~$\LPolishb^\psib$. Given
observations at vertices $v_{b_1}$,~\ldots, $v_{b_C}$, the {\em
  harmonic threat propagation equation\/}
is \begin{equation}\textstyle \Bigl(\LPolishb^\psib_{\rm
    ii}\;\LPolishb^\psib_{\rm ib}\Bigr)\Bigl({\threatb_{\rm i}\atop
    \threatb_{\rm b}}\Bigr) =\zerob\label{eq:hsptpe}\end{equation}
where the generalized Laplacian~$\LPolishb^\psib=
\Bigl({\LPolishb^\psib_{\rm ii}\atop \LPolishb^\psib_{\rm bi}}\,
     {\LPolishb^\psib_{\rm ib}\atop \LPolishb^\psib_{\rm bb}}\Bigr)$
     and the threat vector $\threatb=\Bigl({\threatb_{\rm i}\atop
       \threatb_{\rm b}}\Bigr)$ have been permuted so that observed
     vertices are in the `\/${\rm b}$' blocks (the ``boundary''),
     unobserved vertices are in `\/${\rm i}$' blocks (the
     ``interior''), and the observation vector $\threatb_{\rm b}$ is
     given. The {\em harmonic threat\/} is the solution to
     Eq.~(\ref{eq:hsptpe}), \begin{equation}\threatb_{\rm i} =
       -(\LPolishb^\psib_{\rm ii})^{-1} (\LPolishb^\psib_{\rm
         ib}\threatb_{\rm b}). \label{eq:harmonicthreat}\end{equation}
     Eq.~(\ref{eq:hsptpe}) is directly analogous to Laplace's equation
     ${\Delta\varphi=0}$ given a fixed boundary condition. As
     discussed in the next subsection and
     Section~\ref{sec:connections}, the connection between threat
     propagation and harmonic graph analysis also provides a link to
     spectral-based methods for network detection.  In practice, the
     highly sparse linear system of Eq.~(\ref{eq:harmonicthreat}) may
     be solved by simple repeated iteration of Eq.~(\ref{eq:tp1}), or
     using the biconjugate gradient method, which provides a practical
     computational approach that scales well to graphs with thousands
     of vertices and thousands of time samples in the case of
     space-time threat propagation, resulting in graphs of order
     ten~million or more. In practice, significantly smaller subgraphs
     are encountered in applications such as threat network
     discovery~\cite{Smith2011}, for which linear solvers with sparse
     systems are extremely fast.

\subsubsection{Stochastic Realization Approach}
\label{sec:sptp-stochastic}

The stochastic realization interpretation of the Bayesian threat
propagation equations (\ref{eq:tp1}) is that the
probability of threat for one random walk from~$v$ to the observed
vertex $v_{b_c}$ is \begin{equation} \threat_v\mid\hbox{\rm walk}_{v\to
    v_{b_c}} =\threat_{v_{b_c}},\label{eq:tpwalk}\end{equation} and
the probability of threat~$\threat_v$ at~$v$ equals the threat
probability averaged over all random walks emanating from~$v$.  This
is equivalent to an absorbing Markov chain with absorbing
states~\cite{Pinsky2010} at which random walks terminate. The
absorbing vertices for the threat diffusion model are the $C$ observed
vertices, and an augmented state reachable by all unobserved vertices
representing a transition from threat to non-threat with
probability~${1-\psi_v}$.  The $(N+1)\by(N+1)$ transition matrix for
the Markov chain corresponding to threat propagation equals
\begin{equation} \Tb= \bordermatrix{&\hidewidth\scriptstyle
    N-C\hidewidth&\hidewidth\scriptstyle
    C\hidewidth&\hidewidth\scriptstyle 1\hidewidth\cr
    \hfil\scriptstyle N-C&\Gb&\Hb&\oneb -\psib_{N-C}\cr
    \hfil\scriptstyle C&\zerob&\Ib&\zerob\cr \hfil\scriptstyle
    1&\zerob&\zerob&1\cr} \label{eq:markovthreat} \end{equation} in
which $\Gb$ and~$\Hb$ are defined by the block
partition \begin{equation} \Psib\Db^{-1}\Ab
  =\bordermatrix{&\hidewidth\scriptstyle
    N-C\hidewidth&\hidewidth\scriptstyle C\hidewidth\cr
    \hfil\scriptstyle N-C&\Gb&\Hb\cr \hfil\scriptstyle
    C&*&*\cr} \label{eq:GbHbdef} \end{equation} with `$*$' denoting
unused blocks, and $\psib_{N-C} =(\psi_1,\psi_2,\ldots,\psi_{N-C})^\T$
is the vector of a~priori threat diffusion probabilities from~$1$
to~${N-C}$. The observed vertices $v_{b_1}$,~\dots, $v_{b_C}$ are
assigned to indices ${N-C+1}$,~\dots, $N$, and the augmented
``non-threat'' state is assigned to index~${N+1}$.

According to this stochastic realization model, the threat at a vertex
for any single random walk that terminates at an absorbing vertex is
given by the threat level at the terminal vertex, with the augmented
``non-threat'' vertex assigned a threat level of~zero; the threat is
determined by this result averaged over all random walks.  Ignoring
the a~priori probabilities, this is also precisely the stochastic
realization model for equilibrium thermodynamics and, in general,
solutions to Laplace's equation~\cite{Ozisik1968,Sabelfeld1994}.

As in Eq.~(\ref{eq:Lonekernel}), the uniform vector
$(N+1)^{-1}\oneb_{N+1}$ is the left eigenvector of~$\Tb$ because $\Tb$
is a right stochastic matrix, i.e.\ ${\Tb{\cdot}\oneb=\oneb}$. For an
irreducible transition matrix of a strongly connected graph, the
Perron--Frobenius theorem~\cite{Gantmacher59,Godsil2001} guarantees
that this eigenvalue is simple and that the constant vector is the
unique invariant eigenvector corresponding to~${\lambda=1}$, a trivial
solution that, as usual, poses a fundamental problem for network
detection. However, neither version of the Perron--Frobenius theorem
applies to the transition matrix~$\Tb$ of an absorbing Markov chain
because $\Tb$ is not strictly positive, as required by Perron, nor is
$\Tb$ irreducible, as required by Frobenius---the absorbing states are
not strongly connected to the graph.

To guarantee the existence of nonnegative threat propagating over the
graph, we require a generalization of the Perron--Frobenius theorem for
reducible nonnegative matrices of the form found in
Eq.~(\ref{eq:markovthreat}). The following theorem introduces a new
version of Perron--Frobenius that establishes the existence of a
nonnegative basis for the principal invariant subspace of a reducible
nonnegative matrix.

\begin{theorem} {\bf(Perron--Frobenius for a Reducible Nonnegative
  Matrix).}\quad Let\/ $\Tb$ be a reducible, nonnegative, order~$n$
matrix of canonical form, \begin{equation} \Tb
  = \begin{pmatrix}\Qb&\Rb\\ \zerob&\Ib_r\end{pmatrix},\label{eq:reduciblecanonical}\end{equation}
such that the maximum modulus of the eigenvalues of~$\Qb$ is less than
unity, ${|\lambda_{\rm max}(\Qb)|<1}$, and\/ ${\rank\Rb=r}$. Then the
maximal eigenvalue of\/~$\Tb$ is unity with multiplicity~$r$ and
nondefective. Furthermore, there exists a nonnegative
matrix \begin{equation}\Eb
  =\begin{pmatrix}(\Ib-\Qb)^{-1}\Rb\\ \Ib_r\end{pmatrix}\label{eq:Eb}\end{equation}
of rank~$r$ such
that \begin{equation}\Tb\Eb=\Eb, \label{eq:TE-piss}\end{equation}
i.e.\ the columns of\/~$\Eb$ span the principal invariant subspace
of\/~$\Tb$.
\label{thm:PF-pos-PISS}\end{theorem}

The proof follows immediately by construction and a straightforward
computation involving the partition $\Eb =\Bigl({\Eb_1\atop
  \Eb_2}\Bigr)$ with the choice ${\Eb_2=\Ib_r}$, resulting in the
nonnegative solution to Eq.~(\ref{eq:TE-piss}), $\Eb_1
=(\Ib-\Qb)^{-1}\Rb =(\Ib+\Qb+\Qb^2+\cdots)\Rb$.


Theorem~\ref{thm:PF-pos-PISS} has immediate application to threat
propagation, for by definition the probability of threat on the graph
is determined by the vector $\threatb^{\rm a} =\Bigl({\threatb_{\rm
    i}\atop \threatb^{\rm a}_{\rm b}}\Bigr)$ such that
${\Tb\threatb^{\rm a}=\threatb^{\rm a}}$ and $\threatb^{\rm a}_{\rm
  b}=\Bigl({\threatb_{\rm b}\atop0}\Bigr)$ is determined by the
probabilities of threat $\threatb_{\rm b}
=(\threat_{N-C+1},\ldots,\threat_N)^\T$ at observed vertices
$v_{b_1}$,~\dots, $v_{b_C}$ [cf.\ Eq.~(\ref{eq:hsptpe})] augmented
with zero threat ${\threat^{\rm a}_{N+1}=0}$ at the ``non-threat''
vertex. From Eqs.\ (\ref{eq:markovthreat}) and
(\ref{eq:reduciblecanonical}), ${\Qb=\Gb}$,
$\Rb=\bigl(\Hb\;\;\oneb-\psib_{N-C}\bigr)$, and $\Rb\threatb^{\rm
  a}_{\rm b}=\Hb\threatb_{\rm b}$. Therefore, the vector that
satisfies the proscribed boundary value problem
equals \begin{equation} \threatb^{\rm a}
  =\begin{pmatrix}(\Ib-\Gb)^{-1}\Hb\threatb_{\rm b}\\ \threatb^{\rm
    a}_{\rm b}\end{pmatrix}.\label{eq:PF-threatprop}\end{equation} As
is well-known~\cite{Pinsky2010}, the hitting probabilities of a random
walk from an unobserved vertex to an observed vertex are given by the
matrix ${\Ub=(\Ib-\Gb)^{-1}\Hb}$; therefore, an equivalent definition
of threat probability~$\threat_v$ from Eq.~(\ref{eq:PF-threatprop}) is
the probability that a random walk emanating from~$v$ terminates at an
observed vertex, conditioned on the probability of threat over all
observed vertices: \begin{equation}\threat_v = \sum\nolimits_c
  P(\hbox{\rm walk}_{v\to
    v_{b_c}})P(\Theta_{v_{b_c}}).\label{eq:threat-walk}\end{equation}
We have thus proved the following theorem establishing the equivalence
between the probabilistic and stochastic realization approaches of
threat propagation.

\begin{theorem} {\bf(Harmonic threat propagation).}\quad The vector~$\threatb
=\Bigl({\threatb_{\rm i}\atop \threatb_{\rm b}}\Bigr) \in\Rbbb^N$ is a
solution to the boundary value problem of Eq.~(\ref{eq:hsptpe}) if and
only if the augmented vector $\threatb^{\rm a} =\Bigl({\threatb_{\rm
    i}\atop \threatb^{\rm a}_{\rm b}}\Bigr) \in\Rbbb^{N+1}$ is a
stationary vector of the absorbing Markov chain transition matrix\/
$\Tb$ of Eq.~(\ref{eq:markovthreat}) with given values~$\threatb^{\rm
  a}_{\rm b}$.  Furthermore, $\threatb$ is
nonnegative. \label{thm:prob-stochastic-equal} \end{theorem}

This theorem will also provide a connection to the spectral method for
network detection discussed in Section~\ref{sec:connections}.

\subsection{Space-Time Threat Propagation}
\label{sec:sttp}

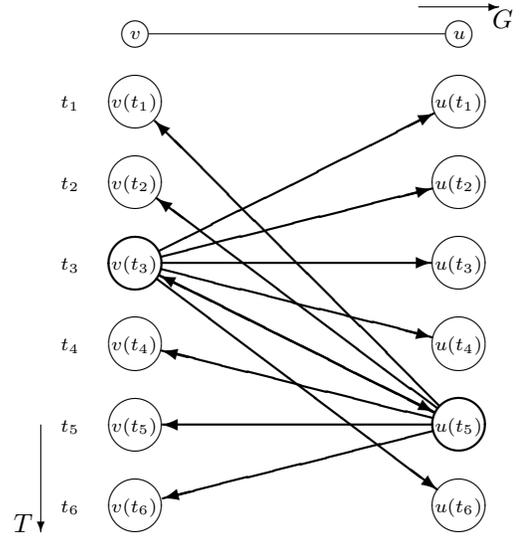
\begin{figure}[t]
\setlength{\unitlength}{0.85pc}
\vskip-0.5\unitlength
\begin{center}
\begin{picture}(18,21)
\put(4,19.5){\circle{1}}
\put(3.8,19.3){$\scriptstyle v$}
\put(16,19.5){\circle{1}}
\put(15.8,19.3){$\scriptstyle u$}
\put(4.5,19.5){\line(1,0){11}}
\put(14.5,20.5){\vector(1,0){3}}
\put(17.25,19.7){$\textstyle G$}
\put(1.25,16.75){$\scriptstyle t_1$}
\put(1.25,13.75){$\scriptstyle t_2$}
\put(1.25,10.75){$\scriptstyle t_3$}
\put(1.25,7.75){$\scriptstyle t_4$}
\put(1.25,4.75){$\scriptstyle t_5$}
\put(1.25,1.75){$\scriptstyle t_6$}
\put(0.5,5){\vector(0,-1){4}}
\put(-0.5,1){$\textstyle T$}
\thinlines
\put(4,17){\circle{2}}
\put(3.25,16.75){$\scriptstyle\! v(t_1)$}
\put(4,14){\circle{2}}
\put(3.25,13.75){$\scriptstyle\! v(t_2)$}
\thicklines
\put(4,11){\circle{2}}
\put(3.25,10.75){$\scriptstyle\! v(t_3)$}
\thinlines
\put(4,8){\circle{2}}
\put(3.25,7.75){$\scriptstyle\! v(t_4)$}
\put(4,5){\circle{2}}
\put(3.25,4.75){$\scriptstyle\! v(t_5)$}
\put(4,2){\circle{2}}
\put(3.25,1.75){$\scriptstyle\! v(t_6)$}
\put(16,17){\circle{2}}
\put(15.25,16.75){$\scriptstyle\! u(t_1)$}
\put(16,14){\circle{2}}
\put(15.25,13.75){$\scriptstyle\! u(t_2)$}
\put(16,11){\circle{2}}
\put(15.25,10.75){$\scriptstyle\! u(t_3)$}
\put(16,8){\circle{2}}
\put(15.25,7.75){$\scriptstyle\! u(t_4)$}
\thicklines
\put(16,5){\circle{2}}
\put(15.25,4.75){$\scriptstyle\! u(t_5)$}
\thinlines
\put(16,2){\circle{2}}
\put(15.25,1.75){$\scriptstyle\! u(t_6)$}
\thicklines
\put(15.29,5.707){\vector(-1,1){10.55}}
\put(15.2,5.6){\vector(-4,3){10.4}}
\put(15.106,5.447){\vector(-2,1){10.2}}
\put(15.03,5.243){\vector(-4,1){10.05}}
\put(15,5){\vector(-1,0){10}}
\put(15.03,4.757){\vector(-4,-1){10}}
\put(4.894,11.447){\vector(2,1){10.2}}
\put(4.97,11.243){\vector(4,1){10.05}}
\put(5,11){\vector(1,0){10}}
\put(4.97,10.757){\vector(4,-1){10.05}}
\put(4.894,10.553){\vector(2,-1){10.2}}
\put(4.8,10.4){\vector(4,-3){10.4}}
\end{picture}
\end{center}
\vskip-1.5\unitlength
\caption{A directed space-time graph~$G_T$ with vertices ${V\times
    T}$, $V=\{\,u,v\,\}$ sampled at index
  times~$T=(t_1,\ldots,t_6)$. For example, an interaction between
  $u(t_5)$ and~$v(t_3)$ (represented by the doubled-sided arrow
  ${v(t_3)\longleftrightarrow u(t_5)}$ above) also creates space-time
  edges via the space-time kernel [Eq.~(\ref{eq:stprob})] between
  $u(t_5)$ and other times at~$v$, and $v(t_3)$ and other times
  at~$u$.\label{fig:stg}}
\end{figure}

Many important network detection applications, especially networks
based on vehicle tracks and computer communication networks, involve
directed graphs in which the edges have departure and arrival times
associated with their initial and terminal vertices. Space-Time threat
propagation is used compute the time-varying threat across a graph
given one or more observations at specific vertices and
times~\cite{Philips2012,Smith2012}. In such scenarios, the
time-stamped graph~${G=(V,E)}$ may be viewed as a {\em space-time
  graph} ${G_T =(V\times T,E_T)}$ where $T$ is the set of sample times
and $E_T\subset[{V\times T}]^2$ is an edge set determined by the
temporal correlations between vertices at specific times. This edge
set is application-dependent, but must satisfy the two constraints,
(1)~if ${\bigl(u(t_k),v(t_l)\bigr)\in E_T}$ then ${(u,v)\in E}$, and
(2)~temporal subgraphs $\bigl((u,v),E_T(u,v)\bigr)$ between any two
vertices $u$ and~$v$ are defined by a temporal model $E_T(u,v)\subset
[{T\sqcup T}]^2$. If the stronger, converse of property~(1) holds,
i.e.\ if ${(u,v)\in E}$ then either ${\bigl(u(t_k),v(t_l)\bigr)\in
  E_T}$ or~${\bigl(v(t_l),u(t_k)\bigr)\in E_T}$ for~all $t_k$, $t_l$,
then if the graph $G$ is irreducible, then so is the space-time graph
$G_T$. An example space-time graph is illustrated in
Fig.~\ref{fig:stg}. The general models for spatial threat propagation
provided in the preceding subsection will now be augmented to include
dynamic models of threat propagation.

Given an observed threat at a particular vertex and time, we wish to
compute the inferred threat across all vertices and all times. Given a
vertex~$v$, denote the threat at~$v$ and at time~${t\in\Rbbb}$ by the
$\{0,1\}$-valued stochastic process $\Theta_v(t)$, with value zero
indicating no threat, and value unity indicating a threat. As above, denote
the\/ {\em probability of threat\/} at~$v$ at~$t$ by
\begin{equation}
  \threat_v(t) \buildrel{\rm def}\over= P\bigl(\Theta_v(t)=1\bigr)
  =P\bigl(\Theta_v(t)\bigr). \label{eq:Pthreat}
\end{equation}
The threat state at~$v$ is modeled by a finite-state continuous time
Markov jump process between from state~$1$ to state~$0$ with Poisson
rate~$\lambda_v$. With this simple model the threat stochastic process
$\Theta_v(t)$ satisfies the It\^o stochastic differential
equation~\cite{Stirzaker2005},
\begin{equation} d\Theta_v =-\Theta_v\,dN_v;\quad
  \Theta_v(0)=\theta_1, \label{eq:sttpIto}
\end{equation}
where $N_v(t)$ is a Poisson process with rate~$\lambda_v$ defined for
positive time, and simple time-reversal provides the model for
negative times.  Given an observed threat ${z=\Theta_v(0)=1}$ at~$v$
at~${t=0}$ so that ${\threat_V(0)=1}$, the probability of threat
at~$v$ under the Poisson process model (including time-reversal) is
\begin{equation} \threat_v(t)
  =P\bigl(\Theta_v(t)\mid z=\Theta_v(0)=1\bigr)
  =e^{-\lambda_v|t|}, \label{eq:stprob}
\end{equation}
This stochastic model provides a Bayesian framework for inferring, or
propagating, threat at a vertex over time given threat at a specific
time. The function \begin{equation}
  K_v(t)=e^{-\lambda_v|t|}\label{eq:stkernel}\end{equation} of
Eq.~(\ref{eq:stprob}) is called the {\em space-time threat kernel\/}
and when combined with spatial propagation provides a temporal model
$E_T$ for a space-time graph. A Bayesian model for propagating threat
from vertex to vertex will provide a full space-time threat
diffusion model and allow for the application of the optimum maximum
likelihood test that will be developed in Section~\ref{sec:npnetdet}.

Propagation of threat from vertex to vertex is determined by
interactions between vertices. Upon computation of the space-time
adjacency matrix, the spatial analysis of Section~\ref{sec:sptp}
applies directly to space-time graphs whose vertices are space-time
positions. The threat at vertex~$v$ at which a single interaction~$\tau$
from vertex~$u$ arrives and\slash or departs at times $t_{\tau}^v$
and~$t_{\tau}^{u}$ is determined by Eq.~(\ref{eq:stprob}) and the
(independent) event $\Psi_v(t)$ that threat propagates through~$v$ at
time~$t$: $P\bigl(\Theta_v(t)\bigr) =\threat_v(t)
=\threat_{u}(t_{\tau}^{u})\* K_v(t-t_{\tau}^v)\* \psi_{v(t)}$. There
is a linear transformation
\begin{multline}\threat_v(t) =\psi_{v(t)}K(t-t_{\tau}^v)\threat_{u}(t_{\tau}^{u})\\ =\int_{-\infty}^\infty
    \psi_{v(t)} K(t-t_{\tau}^v) \delta(\sigma-t_{\tau}^{u})
    \threat_{u}(\sigma)\,d\sigma\end{multline} from the threat
    probability at~$u$ to~$v$. Discretizing time, the temporal matrix
    $\Kb_{\tau}^{uv}$ for the discretized operator has the sparse form
\begin{equation} \Kb_{\tau}^{uv}
    =\Bigl(\>\zerob\;\ldots\;\zerob\>
    K(t_k-t_\tau^v)\>\zerob\;\ldots\;\zerob\>\Bigr),\label{eq:Kuv}
\end{equation} where $\zerob$ represents an all-zero column, $t_k$
represents a vector of discretized time, and the discretized function
$K(t_k-t_{\tau}^v)$ appears in the column corresponding to the
discretized time at~$t_{\tau}^{u}$.  Threat propagating from
vertex~$v$ to~$u$ along the same interaction $\tau$ is given by the
comparable expression $\threat_{u}(t) =\threat_v(t_{\tau}^v)
K(t-t_{\tau}^{u})$, whose discretized linear operator
$\Kb_{\tau}^{vu}$ takes the form \begin{equation}\Kb_{\tau}^{vu}
  =\Bigl(\>\zerob\;\ldots\;\zerob\>
  K(t_k-t_{\tau}^{u})\>\zerob\;\ldots\;\zerob\>\Bigr)\end{equation}
[cf.\ Eq.~(\ref{eq:Kuv})] where the nonzero column corresponds
to~$t_{\tau}^v$. The sparsity of~$\Kb_{\tau}^{uv}$
and~$\Kb_{\tau}^{vu}$ will be essential for practical space-time
threat propagation algorithms. The collection of all interactions
determines a weighted space-time adjacency matrix $\Ab$ for the
space-time graph $G_T$. This is a matrix of order~${\#V{\cdot}\#T}$
whose temporal blocks for interactions between vertices $u$ and~$v$
equals, \begin{equation}
  \begin{pmatrix}\Ab_{uu}&\Ab_{uv}\\ \Ab_{vu}&\Ab_{vv}
  \end{pmatrix} =\begin{pmatrix}\zerob
  &\sum_l\Kb_{\tau_l}^{vu}\\ \sum_l\Kb_{\tau_l}^{uv}
  &\zerob\end{pmatrix}.\label{eq:Ast}\end{equation} Note that with the
space-time threat kernel of~Eq.~(\ref{eq:stkernel}), if~$G$ is
irreducible, then so is~$G_T$.

As with spatial-only threat propagation of Eq.~(\ref{eq:tp1}), the
{\em space-time threat propagation equation\/}
is \begin{align}\threatb
  &=\Psib\Wb^{-1}\Ab\threatb \label{eq:sttpe}\\ \hbox{or}\quad
  \threat_v(t_k) &={\psi_v(t_k)\over \sum\nolimits_{u,l}k_{vu;kl}}
  \sum\nolimits_{u,l}k_{vu;kl}\threat_u(t_l) \label{eq:sttpe-vukl} \end{align}
in which $\threatb$ is the (discretized) space-time vector of threat
probabilities, ${\Ab=(k_{vu;kl})}$ is the (weighted) space-time
adjacency matrix, and ${\Wb=\Diag(\Ab{\cdot}\oneb)}$ and~$\Psib
=\diag\bigl(\psi_1(t_1),\ldots,\psi_N(t_{\#T})\bigr)$ are,
respectively, the space-time diagonal matrices of the space-time
vertex weights and a~priori probabilities that threat propagates
through each spatial vertex at a specific time. In contrast to the
treatment of spatial-only threat propagation in
Section~\ref{sec:sptp}, the space-time graph is necessarily a directed
graph, consistent with the asymmetric space-time adjacency matrix of
Eq.~(\ref{eq:Ast}).

By assumption, the graph $G$ is irreducible, implying that the
space-time graph $G_T$ is also irreducible. Therefore,
Theorem~\ref{thm:maxprincthreatprop} implies a well-defined solution to
the space-time threat propagation equation of Eq.~(\ref{eq:sttpe}) for
a set observations at specific vertices and times,
$v_{b_1}(t_{b_1})$,~\dots, $v_{b_C}(t_{b_C})$. Yet the
Perron--Frobenius theorem for the space-time Laplacian
${\LPolishb=\Ib-\Wb^{-1}\Ab}$ poses precisely the same detection
challenge as with spatial-only propagation: if the a~priori
probabilities are constant and equal to unity, i.e.\ ${\Psib=\Ib}$,
and the observed probability of threat is constant, then the
space-time probability of threat is also constant for~all spatial
vertices and all times, yielding a hopeless detection method.

However, the advantage of time-stamped edges is that the times can be
used to detected temporally coordinated network activity---we seek to
detect vertices whose activity is correlated with that of threat
observed at other vertices. According to this model of threat
networks, the a~priori probability that a threat propagates through
vertex~$v$ at time~$t_k$ is determined by the Poisson process used to
model the probability of threat as a function of
time: \begin{equation} \psi_v(t_k) = {1\over
    d_v}\sum_{u,l}k_{vu;kl}, \label{eq:psi-st-coord}\end{equation}
where $d_v$ is the spatial degree of vertex~$v$, i.e.\ the number of
interactions associated with a spatial vertex. If all interactions arrive\slash
depart at the same time at~$v$, then the a~priori probability of
threat diffusion is unity at this time, but different times reduce
this probability according to the stochastic process for threat. Thus
space-time threat propagation for coordinated activity is determined
by the threat propagation equation, \begin{equation}\threatb
  =\Db^{-1}\Ab\threatb \label{eq:sttpe-coord}\end{equation} in which
$\Db=\diag\bigl(d_1\Ib,\ldots,d_N\Ib\bigr)$ is the block-diagonal
space-time matrix of (unweighted) spatial degrees and $\Ab$ is the
weighted space-time adjacency matrix as in Eq.~(\ref{eq:sttpe}). This
algorithm may also be further generalized to account for spatial-only
a~priori probability models such as the distance from observed
vertices by replacing $1/d_v$ in Eq.~(\ref{eq:sttpe-coord}) with
$\psi'_v/d_v$ and an a~priori model as in Eq.~(\ref{eq:dwtp}),
yielding the threat propagation equation
${\threatb=\Psib'\Db^{-1}\Ab\threatb}$ with
$\Psib'=\diag\bigl(\psi'_1\Ib,\ldots,\psi'_N\Ib\bigr)$.

\subsection{Hybrid Threat Propagation}
\label{sec:hybridtp}

The temporal kernels introduced for time-stamped edges in
Section~\ref{sec:sttp} are appropriate for network detection
applications that involve time-stamped edges; however, there are many
applications in which such time-stamped information is either
unavailable, irrelevant, or uncertain. Ignoring small routing delays,
computer network communication protocols occur essentially
instantaneously, and text documents may describe relationships between
sites independent of a specific timeframe. Integrating spatio-temporal
relationships from multiple information sources necessitates a hybrid
approach combining, where appropriate, the spatial-only capabilities
of Section~\ref{sec:sptp} with the space-time methods of
Section~\ref{sec:sttp}.

In situations such as computer communication networks in which the
timescale of the relationship is much smaller than the discretized
timescale, then connections from one vertex to another arrive at the
same discretized time, and the temporal blocks for connections between
vertices $u$ and~$v$ replaces Eq.~(\ref{eq:Ast}) and
equals, \begin{equation}\begin{pmatrix}\Ab_{uu}&\Ab_{uv}\\ \Ab_{vu}&\Ab_{vv}
  \end{pmatrix} =\begin{pmatrix}\zerob
  &\Ib\\ \Ib &\zerob\end{pmatrix}.\label{eq:Ast-I}\end{equation} In
situations such as time-independent references within text documents
in which threat at any time at vertex~$u$ implies a threat at all
times at vertex~$v$, and vice~versa, the temporal blocks for
connections between vertices $u$ and~$v$
equals, \begin{equation}\begin{pmatrix}\Ab_{uu}&\Ab_{uv}\\ \Ab_{vu}&\Ab_{vv}
  \end{pmatrix} =\begin{pmatrix}\zerob
  &(\#T)^{-1}\oneb\oneb^\T\\ (\#T)^{-1}\oneb\oneb^\T
  &\zerob\end{pmatrix},\label{eq:Ast-clique}\end{equation} i.e.\ a
space-time clique between $u$ and~$v$. This equivalent the space-time
model with Poisson rate ${\lambda=0}$.

\subsection{Neyman--Pearson Network Detection}
\label{sec:npnetdet}

Network detection of a subgraph within a graph ${G=(V,E)}$ of
order~$N$ is treated as $N$ independent binary hypothesis tests to
decide which of the graph's $N$ vertices do not belong (null
hypothesis $H_0$) or belong (hypothesis $H_1$) to the
network. Maximizing the probability of detection (PD) for a fixed
probability of false alarm (PFA) yields the Neyman--Pearson test
involving the log-likelihood ratio of the competing hypotheses. We
will derive this test in the context of network detection, which both
illustrates the assumptions that ensure detection optimality, as well
as indicates practical methods for computing the log-likelihood ratio
test and achieving an optimal network detection algorithm. It will be
seen that a few basic assumptions yield an optimum test that is
equivalent to the Bayesian threat propagation algorithm developed in
the previous section.  If any part of the graph is unknown or
uncertain, then the Markov transition probabilities may be treated as
random variables and either marginalized out of the likelihood ratio,
yielding Neyman--Pearson optimality in the average sense, or the
maximum likelihood estimate may be used in the suboptimum generalized
likelihood ratio test (GLRT)~\cite{VanTrees1968}. We will not cover
extensions to unknown parameters in this paper. The optimum test
involves the graph Laplacian, which allows comparison of
Neyman--Pearson testing to several other network detection methods
whose algorithms are also related to the properties of the Laplacian.

An optimum hypothesis test is now derived for the presence of a
network given a set of observations $\zb$ according to the observation
model of Definition~\ref{def:om}.  Optimality is defined in the
Neyman--Pearson sense in which the probability of detection is
maximized at a constant false alarm rate
(CFAR)~\cite{VanTrees1968}. For the general problem of network
detection of a subgraph within graph $G$ of order~$N$, the decision of
which of the $2^N$ hypothesis
${\Thetab=(\Theta_{v_1},\ldots,\Theta_{v_N})^\T}$ to choose involves a
\hbox{$2^N$-ary} multiple hypothesis test over the measurement space
of the observation vector~$\zb$, and an optimal test involves
partitioning the measurement space into $2^N$ regions yielding a
maximum PD. This NP-hard general combinatoric problem is clearly
computationally and analytically intractable. However,
Eq.~(\ref{eq:Threat-v-walk}) following Definition~\ref{def:threatprop}
guarantees that the threats at each vertex are independent random
variables, allowing the general \hbox{$2^N$-ary} multiple hypothesis
test to be greatly simplified by treating it as $N$ independent binary
hypothesis tests at each vertex.

At each vertex ${v\in G}$ and unknown threat $\Theta\colon
V\to\{0,1\}$ across the graph , consider the binary hypothesis test
for the unknown value $\Theta_v$, \begin{equation}\Heqalign{{\rm
    H}_0(v)& \Theta_v=0&&(vertex belongs to background)\cr {\rm H}_1(v)&
    \Theta_v=1&&(vertex belongs to
    foreground).\cr}\label{eq:graphhyp} \end{equation} Given the
observation vector $\zb\colon \{v_{b_1},\ldots,v_{b_C}\}\subset V\to
M\subset\Rbbb^C$ with observation models
$f\bigl(z(v_{b_j})\mid\Theta_{v_{b_j}}\bigr)$, ${j=1}$,~\dots, $C$, the
PD and PFA are given by the integrals $\hbox{PD}= \int_R
f(\zb\mid{\Theta_v=1})\,d\zb$ and $\hbox{PFA}= \int_R
f(\zb\mid{\Theta_v=0})\,d\zb$, where $R\subset M$ is the detection region
in which observations are declared to yield the decision
${\Theta_v=1}$, otherwise $\Theta_v$ is declared to equal~$0$. The
optimum Neyman--Pearson test uses the detection region~$R$ that
maximizes PD at a fixed CFAR value $\PFA_0$, yielding the {\em
  likelihood ratio\/} (LR) test~\cite{VanTrees1968}, \begin{equation}
  \frac{f(\zb\mid{\Theta_v=1})}
       {f(\zb\mid{\Theta_v=0})}\detthreshv{v}\lambda \label{eq:LR}\end{equation}
for some ${\lambda>0}$. Likelihood ratio tests are also used for graph
classification~\cite{Ligo13}.

Finally, a simple application of Bayes' theorem to the harmonic threat
${\threat_v=f(\Threat_v\mid\zb)}$ provides the optimum Neyman--Pearson
detector [Eq.~(\ref{eq:LR})] because \begin{multline}
  \frac{f(\zb\mid{\Threat_v=1})}{f(\zb\mid{\Threat_v=0})}
  =\frac{f({\Threat_v=1}\mid\zb)}{f({\Threat_v=0}\mid\zb)}
  \cdot\frac{f({\Threat_v=0})}{f({\Threat_v=1})}\\ =\frac{\threat_v}{1-\threat_v}
  \cdot\frac{f({\Threat_v=1})}{f({\Threat_v=0})}
  \detthreshv{v}\lambda, \label{eq:LRsttp} \end{multline} results in a
threshold of the harmonic space-time threat propagation vector of
Eq.~(\ref{eq:varthetav}), \begin{equation}\threat_v
  \detthreshv{v}\mbox{threshold}, \label{eq:thetavthresh}\end{equation}
with the prior ratio $f({\Threat_v=1})/f({\Threat_v=0})$ and the
monotonic function ${\threat_v\mapsto \threat_v/(1-\threat_v)}$ being
absorbed into the detection threshold. By construction, the event
${\Theta_v=1}$ is equivalent to a random walk between~$v$ and one of
the observed vertices $v_{b_1}$, \ldots,~$v_{b_C}$, along with one of
the events ${\Theta_{v_{b_1}}=1}$, \ldots, ${\Theta_{v_{b_C}}=1}$, as
represented in Eq.~(\ref{eq:threat-walk}). Note that $\theta_v$ and
equivalently the likelihood ratio are continuous functions of the
probabilities $P(\hbox{\rm walk}_{v\to v_{b_c}})$
and~$P(\Theta_{v_{b_c}})$; therefore, equality of the likelihood ratio
to any given threshold exists only on a set of measure zero.  If the
prior ratio is constant for~all vertices, then the threshold is also
constant, and the likelihood ratio test [Eq.~(\ref{eq:thetavthresh})]
for optimum network detection becomes \begin{equation}\threatb
  \detthresh\mbox{threshold}. \label{eq:sttpthresh}\end{equation} This
establishes the detection optimality of harmonic space-time threat
propagation.

Because the probability of detecting threat is maximized at each
vertex, the probability of detection for the entire subgraph is also
maximized, yielding an optimum Neyman--Pearson test under the
simplification of treating the $2^N$-ary multiple hypothesis testing
problem as a sequence of $N$ binary hypothesis tests. Summarizing, the
probability of network detection given an observation~$\zb$ is
maximized by computing $f(\Threat_v\mid\zb)$ using a Bayesian threat
propagation method and applying a simple likelihood ratio test,
yielding the following theorem that equates threat propagation with
the optimum Neyman--Pearson test.

\begin{theorem} {\bf(Neyman--Pearson Optimality of Threat
  Propagation).}\quad The solution to Bayesian threat propagation
  expressed in Eqs.\ (\ref{eq:hsptpe}) or~(\ref{eq:PF-threatprop})
  yields an optimum likelihood ratio test in the Neyman--Pearson
  sense. \label{thm:tp-optimum}
\end{theorem}

\section{Modeling and Performance}
\label{sec:modeling}

Evaluation of network detection algorithms may be approached from the
perspectives of theoretical analysis or empirical experimentation.
Theoretical performance bounds have only been accomplished for simple
network models,
i.e.\ cliques~\cite{Fortunato2007,Kumpula2007,Nadakuditi2012} or dense
subgraphs~\cite{Arias-Castro2013} embedded within
Erd\humlaut{o}s--R\'enyi backgrounds, and there are no theoretical
results at~all for more complex network models that characterize
real-world networks~\cite{Smith2013}. If representative network data
with truth is available, one may evaluate algorithm performance with
specific data sets~\cite{Yee2012}. However, real-world data sets of
covert networks with truth is unknown to the authors.  Therefore,
network detection performance evaluation must be conducted on
simulated networks using generative models.  We begin with a simple
stochastic blockmodel~\cite{Wasserman1994}, explore this model's
limitations, then introduce a new network model designed to address
these defects while at the same time encompassing the characteristics
of real-world
networks~\cite{Aiello2001,Airoldi2008,Chakrabarti2004,Newman2003,Xu2008}. Varying
model parameters also yields insight on the dependence of algorithm
performance on different network characteristics.

For each evaluation, we compare performance between the space-time
threat propagation [STTP; Section~\ref{sec:sttp}], breadth-first
search spatial-only threat propagation [BFS; Eq.~(\ref{eq:psi-bfs})],
and modularity-based spectral detection algorithm
[SPEC]~\cite{Miller2010b}. The performance metric is the standard
receiver operating characteristic (ROC), which in the case of network
detection is the probability of detection (i.e.\ the percentage of
true foreground vertices detected) versus the probability of false alarms
(i.e.\ the percentage of background vertices detected) as the detection
threshold is varied.

\subsection{Detection Performance On Stochastic Blockmodels}
\label{sec:StochBlockmodel}

\subsubsection{Stochastic Blockmodel Description}
\label{sec:StochBlockmodelDescription}

The stochastic blockmodel captures the sparsity of real-world networks
and basic community structure~\cite{Girvan02} using a simple network
framework~\cite{Wasserman1994}. For a graph of order~$N$ divided into
$K$ communities, the model is parameterized by a $N\by K$ $\{0,1\}$
membership matrix $\Pib$ and a $K\by K$ probability matrix $\Sb$ that
defines the probability of an edge between two vertices based upon
their community membership. Therefore, the probability of an edge is
determined by the off-diagonal terms of the matrix
$\Pib\Sb\Pib^\T$. By the classical result of
Erd\humlaut{o}s--R\'enyi~\cite{Erdos1960}, each community is almost
surely connected if $\Sb_{kk}>\log N_k/N_k$ in which $N_k$ is the
number of vertices in community~$k$. We introduce the activity
parameter ${r_k\ge1}$ and set $\Sb_{kk}=r_k\log N_k/N_k$ to adjust a
community's density relative to its Erd\humlaut{o}s--R\'enyi
connectivity threshold.

\subsubsection{Experimental Setup and Results}
\label{sec:StochBlockmodelExp}

\begin{figure}[t]
\medskip
\normalsize
\centerline{\includegraphics[width=0.8\linewidth]{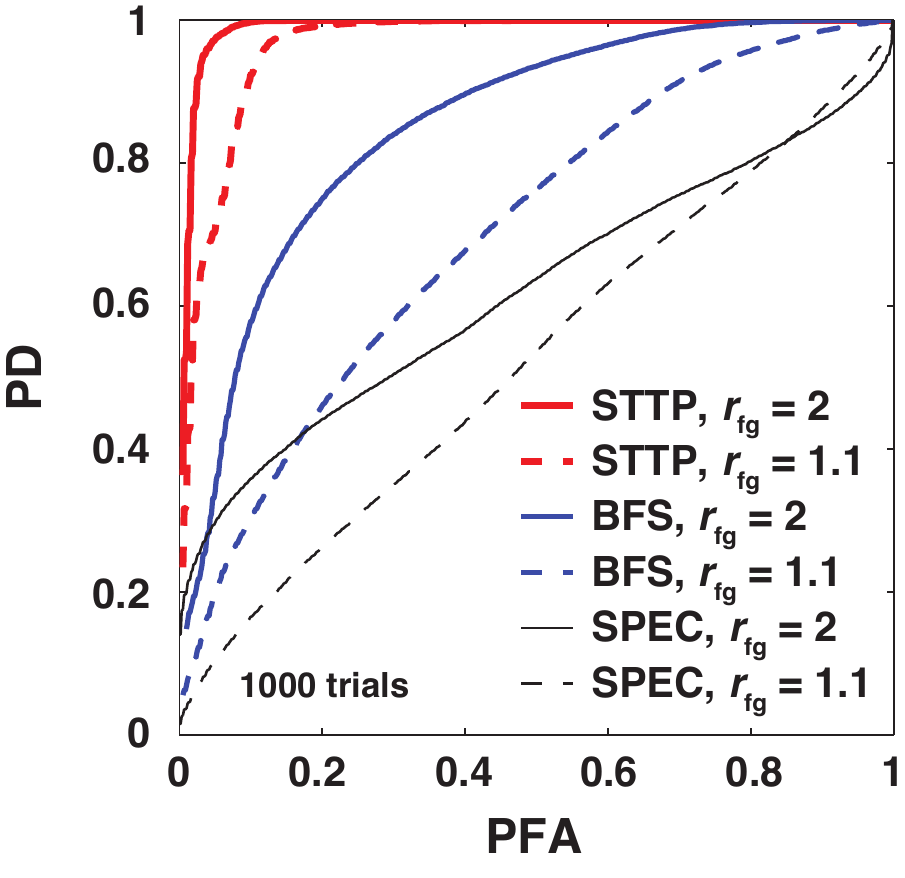}}
\caption{Detection ROC curves of the three different algorithms at two
  levels of foreground activity, $1.1{\cdot}\log N_{\rm fg}/N_{\rm
    fg}$ and $2{\cdot}\log N_{\rm fg}/N_{\rm fg}$. Data is simulated
  using the stochastic blockmodel with 1000 Monte~Carlo trials each
  with an independent draw of the random network and single threat
  observation.\label{fig:ERplot}}
\end{figure}

The objective of this experiment is to quantify detection performance
of a foreground network with varying activity given observations from
a small fraction of its members.  Fig.~\ref{fig:ERplot}
illustrates the ROC detection performance with a graph of order
${N=256}$ and ${K=3}$ with two background communities of order~$128$,
and a foreground community of order~$30$ randomly embedded in the
background. The probability matrix is, \begin{equation}\Sb
  =\begin{pmatrix}0.08&0.02&0.02\\ 0.02&0.08&0.02\\ 0.02&0.02&r_{\rm
    fg}{\cdot}0.1 \nonumber\end{pmatrix},\end{equation} parameterized
by the activity $r_{\rm fg}$ relative to the Erd\humlaut{o}s--R\'enyi
foreground connectivity threshold ${\log30/30\approx0.1}$. A simple
temporal model is used with all foreground interactions at the same
time (i.e.\ perfect coordination), and background interactions
uniformly distributed in time (i.e.\ uncoordinated).

Results are shown for both sparsely connected (${r_{\rm fg}=1.1}$) and
moderately connected (${r_{\rm fg}=2}$) foreground networks. The
simulations show that excellent ROC performance is achievable if
temporal information is exploited (STTP) with highly coordinated
foreground networks with sparse to moderate connectivity. Because of
the use of temporal information, STTP outperforms BFS. Spectral
methods, which are designed to detect highly connected networks
perform poorly on sparse foreground networks, and improve as
foreground network connectivity increases, especially in the low PFA
region in which SPEC performs better than BFS threat propagation.
This result is consistent with expectations and recent theoretical
results for spectral methods applied to clique
detection~\cite{Nadakuditi2012, Arias-Castro2013}.  Continuous
likelihood ratio tests possess ROC curves that are necessarily convex
upwards~\cite{VanTrees1968}; therefore, the ROCs for threat
propagation algorithms applied to data generated from random walk
propagation are necessarily convex. The results of
Fig.~\ref{fig:ERplot} show both threat propagation and spectral
methods applied to data generated from a stochastic
blockmodel. Because the spectral detection algorithm is not associated
with a likelihood ratio test, convexity of its ROC curves is not
guaranteed---indeed, the spectral ROC curve with ${r_{\rm fg}=2}$ is
seen to be concave in the high PD region. All threat propagation ROC
curves are observed to be convex, except for a small part of STTP with
${r_{\rm fg}=1.1}$ near ${\hbox{PD}=0.7}$. This slight concavity
(about $2\%$) may be caused by model mismatch between the stochastic
blockmodel and the random walk model, or statistical fluctuation of
the Monte~Carlo analysis (about $1.4\%$ binomial distribution variance
at ${\hbox{PD}=0.7}$).

Of course, real-world networks are not perfectly coordinated, ideal
Erd\humlaut{o}s--R\'enyi graphs. We will develop a novel, more
realistic model in the next section to explore how more realistic
networks with varying levels of foreground coordination affect the
performance of space-time threat propagation.

\subsection{Detection Performance on the Hybrid Mixed-Membership Blockmodel}
\label{sec:complexBlockmodel}

\subsubsection{Hybrid Mixed-Membership Blockmodel Description}
\label{sec:HybridBlockmodelDescription}

\begin{figure}[t]
\medskip
\normalsize
\centerline{\includegraphics[width=\linewidth]{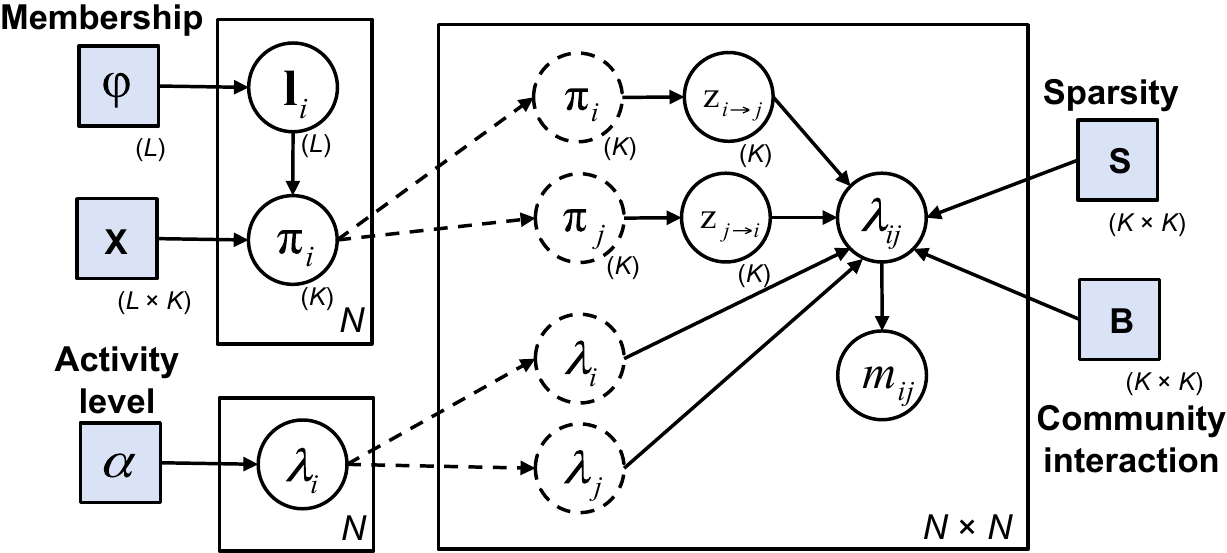}}
\caption{Hybrid mixed-membership blockmodel for the network simulation with $N$
  vertices, $K$ communities, and $L$ lifestyles. Shaded squares are
  model parameters for tuning and circles are variables drawn during
  simulation.\label{fig:blockmodel}}
\end{figure}

Real-world networks display basic topological characteristics that
include a power-law degree distribution (i.e.\ the ``small world''
property)~\cite{Chakrabarti2004}, mixed-membership-based community
structure (i.e.\ individuals belong to multiple
communities)~\cite{Wasserman1994,Airoldi2008}, and
sparsity~\cite{Newman2006}. No one simple network model captures all
these traits. For example, the stochastic blockmodel above provides
sparsity and a rough community structure, but does not capture
interactions through time, the power-law degree distribution, nor the
reality that each individual may belong to multiple communities. The
power-law models such as R-MAT~\cite{Chakrabarti2004} do not capture
membership-based community structure, and mixed-membership stochastic
blockmodels~\cite{Airoldi2008} does not capture power-law degree
distribution nor temporal coordination. To capture all these
characteristics of the real-world networks model, we propose a new
parameterized generative model called the ``hybrid mixed-membership
blockmodel'' (HMMB) that combines the features of these fundamental
network models.  The proposed model is depicted as the plate diagram
in Fig.~\ref{fig:blockmodel}.

The hybrid mixed-membership blockmodel is an aggregate of the
following simpler models and their features: Erd\humlaut{o}s--R\'enyi
for sparsity~\cite{Erdos1960}, Chung--Lu for power-law degree
distribution~\cite{Aiello2001}, and mixed-membership stochastic
blockmodel for community structure~\cite{Airoldi2008}. We model the
number of interactions between any two individuals (i.e.\ edge
weights) as Poisson random variables. Each interaction receives a
timestamp through a coordination model. As above, let $N$ be the order
of the graph, and let $K$ be the number of communities. Each
individual (i.e.\ vertex) divides its membership among the $K$
communities (i.e.\ mixed membership), and the fraction in which an
individual participates among the different communities is determined
by $L$ distinct lifestyles.  The rate $\lambda_{ij}$ of interactions
between vertices $i$ and~$j$ is given by the product
\begin{equation}\lambda_{ij} =I_{ij}^\Sb\cdot
  \frac{\lambda_i\lambda_j} {\sum_k\lambda_k}\cdot\zb_{i\rightarrow
    j}^\T \Bb \zb_{j\rightarrow i},\label{eq:bmrates}\end{equation}
where the first term $I_{ij}^\Sb$ is the (binary) indicator function
drawn from the stochastic blockmodel described in
Section~\ref{sec:StochBlockmodel}, the second term
$\lambda_i\lambda_j/\bigl(\sum_k\lambda_k\bigr)$ is the Chung--Lu model
with per-vertex expected degrees~$\lambda_i$, and the third term
$\zb_{i\rightarrow j}^\T\Bb\zb_{j\rightarrow i}$ is the
mixed-membership stochastic blockmodel with $K\by K$ block matrix
$\Bb$ that determines the intercommunity interaction strength, and
$\zb_{i\rightarrow j}$ is a $\{0,1\}$-valued $K$-vector that indicates
which community membership that vertex~$i$ assumes when interacting
with vertex~$j$.

The mixed-membership $K$-vector $\pib_i$ specifies the fraction that
individual vertex~$i$ divides its membership among the $K$ communities
so that ${\oneb^\T\pib_i\equiv1}$. Each vertex is assigned, via the
$\{0,1\}$-valued $L$-vector $\lb_i$, to one of~$L$ ``lifestyles'' each
with an expected membership distribution given by the $L\by K$ matrix
$\Xb$. The membership distribution $\pib_i$ is determined by a
Dirichlet random draw using the $K$-vector $\lb^\T\Xb$. The lifestyle
vector $\lb_i$ is determined from a multinomial random draw using the
$L$-vector $\phib$ as the probability of belonging to each
lifestyle. Similarly, for~each interaction, the community indicator
vector $\zb_{i\rightarrow j}$ is determined from a multinomial random
draw using the $K$-vector $\pib_i$ as the probability of belonging to
each community. The expected vertex degrees $\lambda_i$ are determined
from a power-law random draw using the exponent~$\alpha$.  The
parameter matrices $\Sb$ and~$\Bb$ are fixed.

Finally, intracommunity coordination is achieved by the nonnegative
$K$-vector $\gammab$, a Poisson parameter of the average number of
coordinated events at each vertex within a specific community. Smaller
values of~$\gammab_k$ correspond to higher levels of coordination in
community~$k$ because there are fewer event times from which to
choose. A community-dependent Poisson random draw determines the
integer number of event times within each community, which are then
drawn uniformly over the time extent of interest. An edge between
vertices $i$ and~$j$ is assigned two random event timestamps based on
the community indicator vectors $\zb_{i\rightarrow j}$ and
$\zb_{j\rightarrow i}$.

\begin{figure}[t]
\medskip
\normalsize
\centerline{\includegraphics[width=0.8\linewidth]{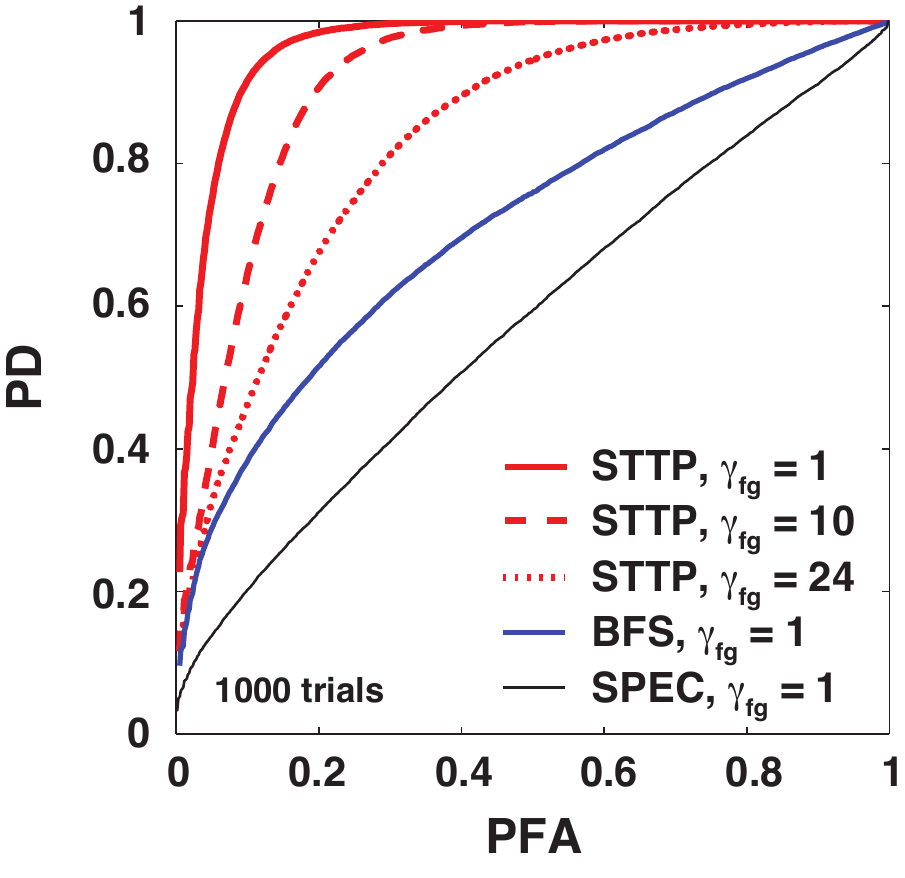}}
\caption{Detection ROC curves of the three different algorithms at
  three levels (${\gamma_{\rm fg}=1}$, $10$, $24$) of foreground
  coordination. Spectral detection and BFS threat propagation do not
  use temporal information so their performance is unaffected by the
  coordination level.\label{fig:complexCoorplot}}
\end{figure}

\subsubsection{Experimental Setup and Results}
\label{sec:HybridBlockmodelExperiment}

The objective of this experiment is to quantify detection performance
with varying coordination of a realistic foreground network operating
within a realistic background.  We use eleven ``lifestyles'' spanning
ten communities, with two lifestyles designated as foreground and all
others as background.

The foreground network's coordination varies from ${\gamma_{\rm
    fg}=1}$ (i.e.\ highly coordinated activity at a single time,
consistent with the tactic used by covert networks to mitigate their
exposure to discovery) to ${\gamma_{\rm fg}=24}$ (i.e.\ less
coordination). Each member of the covert foreground network is also a
member of several background communities.  The foreground and
background order are the same as in the experiment of
Section~\ref{sec:StochBlockmodelExp}, and sparsity levels all ${\log
  N_i/N_i}$.  The foreground network is only a small fraction of the
entire population. Foreground actors are characterized by two distinct
lifestyles representing their memberships in the covert community as
well as different background communities.  The background communities
are intended to represent various business, home, industry, religious,
sports, or other social interactions.

Fig.~\ref{fig:complexCoorplot} illustrates the ROC performance with
these parameters, varying the level of foreground
coordination. Through Eq.~(\ref{eq:psi-st-coord}), space-time threat
propagation is designed to perform well with highly coordinated
networks, consistent with the results observed in
Fig.~\ref{fig:complexCoorplot} in which STTP performs best at the
higher coordination levels and outperforms the breadth-first search
and modularity-based spectral detection methods. The spectral
detection algorithm is expected to perform poorly in this scenario
because, as discussed in Section~\ref{sec:connections}, it relies upon
a relatively dense foreground network, which does not exist in this
simulated dataset with realistic properties of covert networks.

\section{Conclusions}
\label{sec:conc}

A Bayesian framework for network detection can be used to unify the
different approaches of network detection algorithms based on random
walks\slash diffusion and algorithms based on spectral properties.
Indeed, using the concise assumptions for random walks and threat
propagation laid out in Definition~\ref{def:threatprop}, all the
theoretical results follow immediately, including the proof of
equivalence, an exact, closed-form, efficient solution, and
Neyman--Pearson optimality. Not only is this theoretically appealing,
but it provides direct practical benefits through a new network
detection algorithm called space-time threat propagation, that is
shown to achieve superior performance with simulated covert
networks. Bayesian space-time threat propagation is interpreted both
as a random walk on a graph and, equivalently, as the solution to a
harmonic boundary value problem. Bayes' rule determines the unknown
probability of threat on the uncued nodes---the ``interior''---based
on threat observations at cue nodes---the ``boundary.''  Hybrid threat
propagation algorithms appropriate for heterogeneous spatio-temporal
relationships can be obtained from this general threat diffusion
model.  This new method is compared to well-known spectral methods by
examining competing notions of network detection optimality.  To model
realistic covert networks realistically embedded within realistic
backgrounds, a new hybrid mixed-membership blockmodel based on mixed membership
of random graphs is introduced and used to assess algorithm detection
performance on graphs with varying activity and coordination.  In the
important situations of low foreground activity with varying levels of
coordination, the examples show the superior detection performance of
Bayesian space-time threat propagation compared to other spatial-only
and uncued spectral methods.


\section*{Acknowledgments}

The authors gratefully acknowledge the consistently incisive and
constructive comments from our reviewers, which greatly improved this
paper. We also thank Professor Patrick Wolfe for originally suggesting
the deep connection with random walks on graphs.

\let\bibliographysize=\footnotesize 

\bibliography{strings,refs}

\endgroup 


\begin{IEEEbiography}[{\includegraphics[width=1in,clip,keepaspectratio]{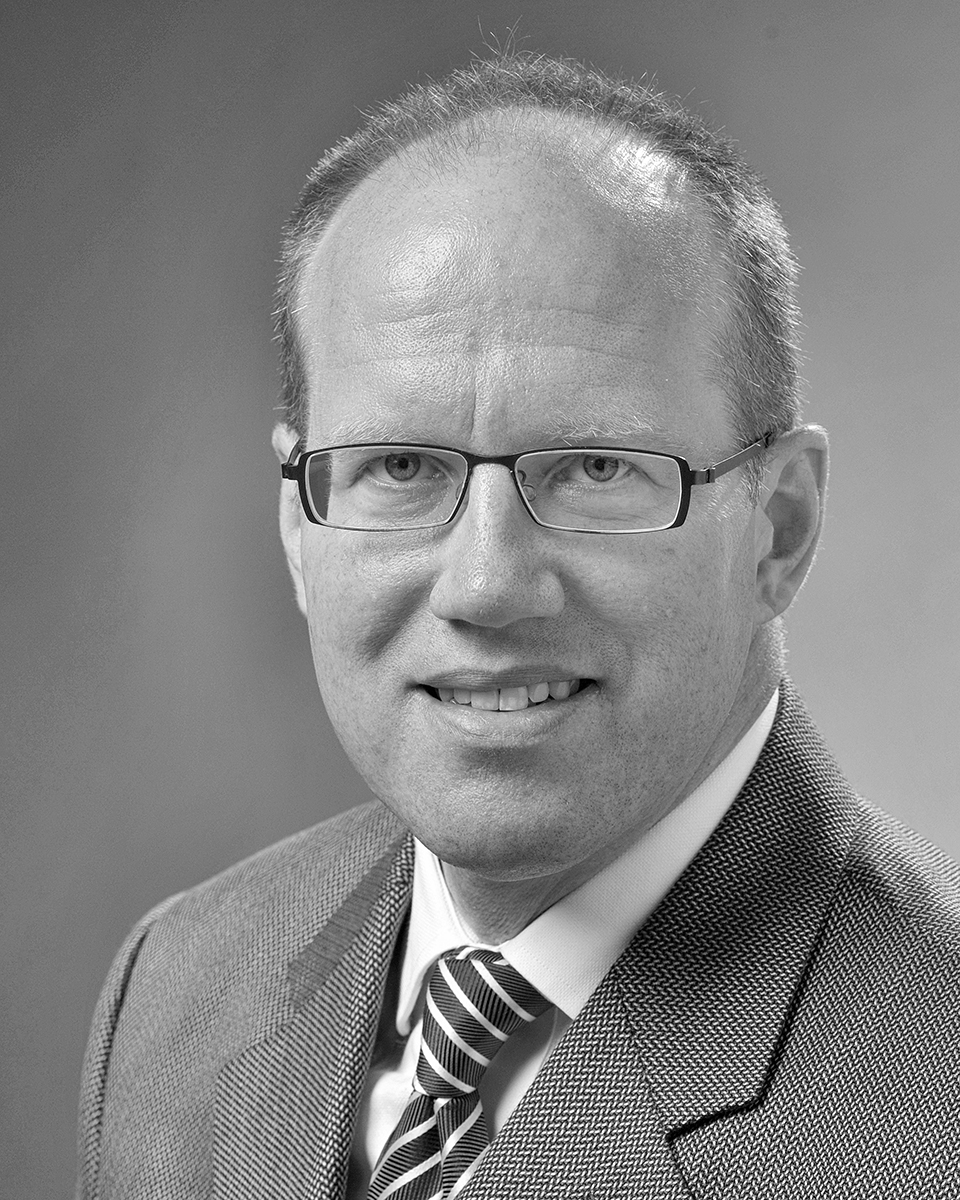}}]{Steven
  Thomas Smith} (M'86--SM'04) is a Senior Staff Member at MIT Lincoln
Laboratory, Lexington,~MA.  He received the B.A.Sc.\ degree in
electrical engineering and mathematics from the University of British
Columbia, Vancouver,~BC in 1986 and the Ph.D. degree in applied
mathematics from Harvard University, Cambridge,~MA in 1993. He has
over 15 years experience as an innovative technology leader with
statistical data analytics, both theory and practice, and broad
leadership experience ranging from first-of-a-kind algorithm
development for groundbreaking sensor systems to graph-based
intelligence architectures. His contributions span diverse
applications from optimum network detection, geometric optimization,
geometric acoustics, statistical resolution limits, and nonlinear
parameter estimation. He received the SIAM Outstanding Paper Award in
2001 and the IEEE Signal Processing Society Best Paper Award in 2010.
He was associate editor of the {\it IEEE Transactions on Signal
  Processing\/} in 2000--2002, and currently serves on the IEEE Sensor
Array and Multichannel committee.  He has taught signal processing
courses at Harvard and for the IEEE.
\end{IEEEbiography}

\begin{IEEEbiography}[{\includegraphics[width=1in,clip,keepaspectratio]{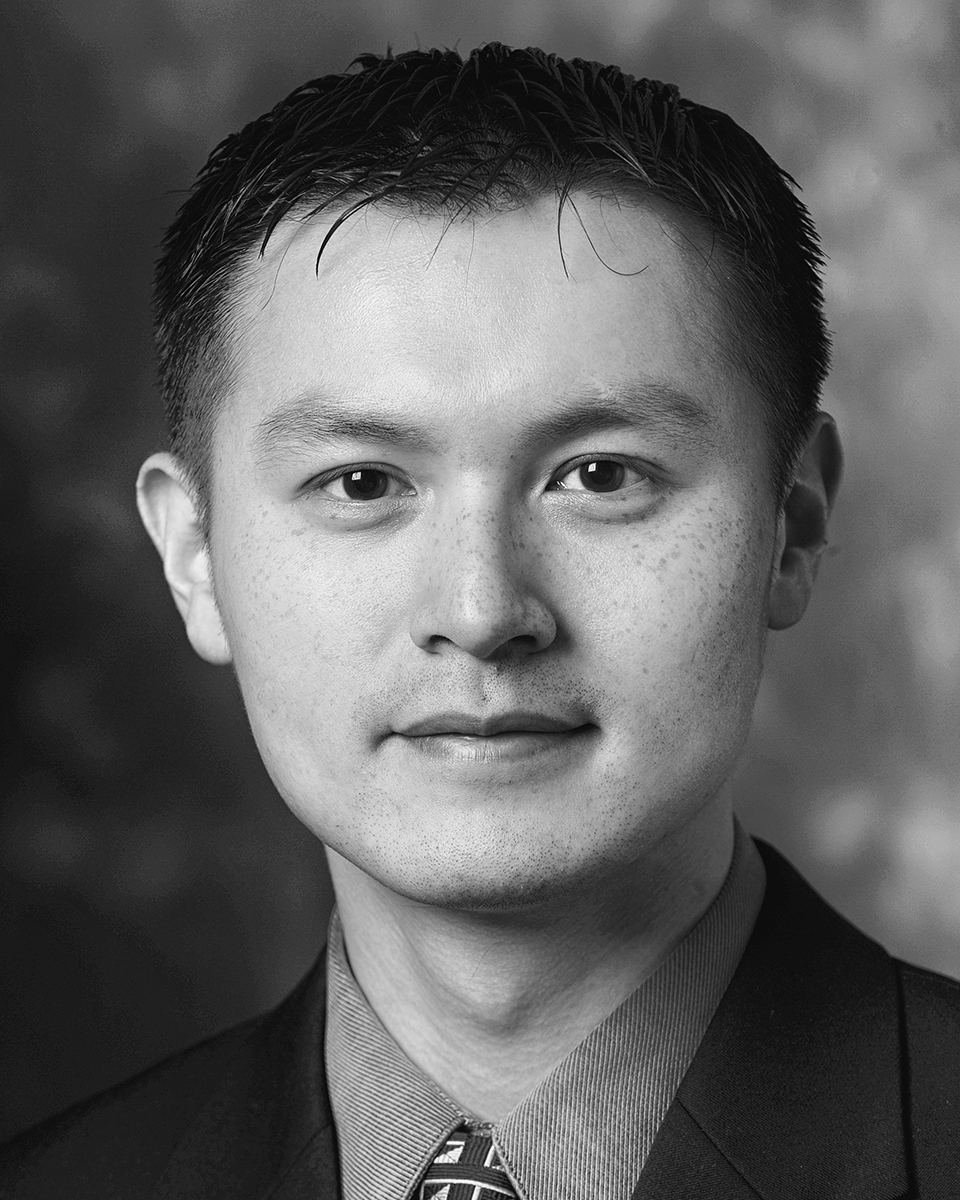}}]{Edward
  K. Kao} (M'03)
is a Lincoln scholar at MIT Lincoln Laboratory in the Intelligence and
Decision Technologies Group. Since joining Lincoln in 2008, he has
been working on graph-based intelligence, where actionable
intelligence is inferred from interactions and relationships between
entities. Applications include wide area surveillance, threat network
detection, homeland security, and cyber warfare, etc. In 2011, he
entered the Ph.D. program at Harvard Statistics. Current research
topics include: causal inference on peer influence effects,
statistical models for community membership estimation, information
content in network inference, and optimal sampling and experimental
design for network inference.
\end{IEEEbiography}

\begin{IEEEbiography}[{\includegraphics[width=1in,clip,keepaspectratio]{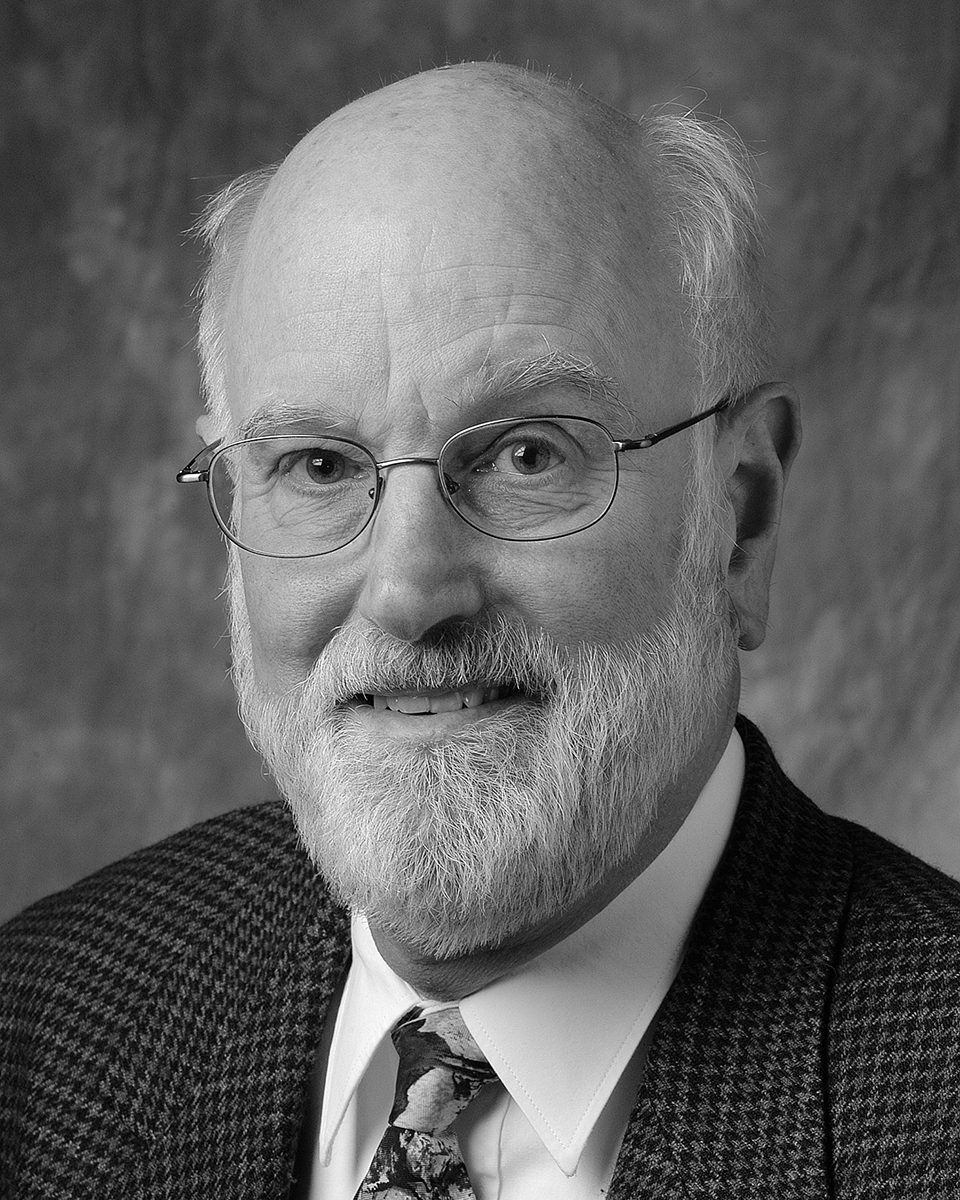}}]{Kenneth~D. Senne}
(S'65--M'72--SM'95--F'02--LF'08) serves as Principal Staff in the ISR
and Tactical Systems Division at Lincoln Laboratory.  His research
examines the application of large data analytics to decision support
problems.  He joined the Laboratory in 1972 to work on the design and
collision avoidance application of the Mode-S beacon system for the
Federal Aviation Administration.  From 1977 to 1986 he contributed to
the development of anti-jam airborne communication systems and super
resolution direction finding with adaptive antennas.  In 1986 he was
asked to set up an array signal processing group as part of a large
air defense airborne electronics program.  This effort resulted in the
pioneering demonstration of a large scale, real-time embedded adaptive
signal processor. In 1998 he was promoted to head the Air Defense
Technology Division. In 2002 he established the Laboratory's
Technology Office, with responsibility for managing technology
investments, including the internal innovative research program.
Prior to joining Lincoln Laboratory he earned a Ph.D. degree in
electrical engineering from Stanford University with foundational
research on digital adaptive signal processing and he served as
Captain in the U.S. Air Force with the Frank J. Seiler Research
Laboratory at the Air Force Academy.  He was elected Fellow of the
IEEE in 2002.
\end{IEEEbiography}

\begin{IEEEbiography}[{\includegraphics[width=1in,clip,keepaspectratio]{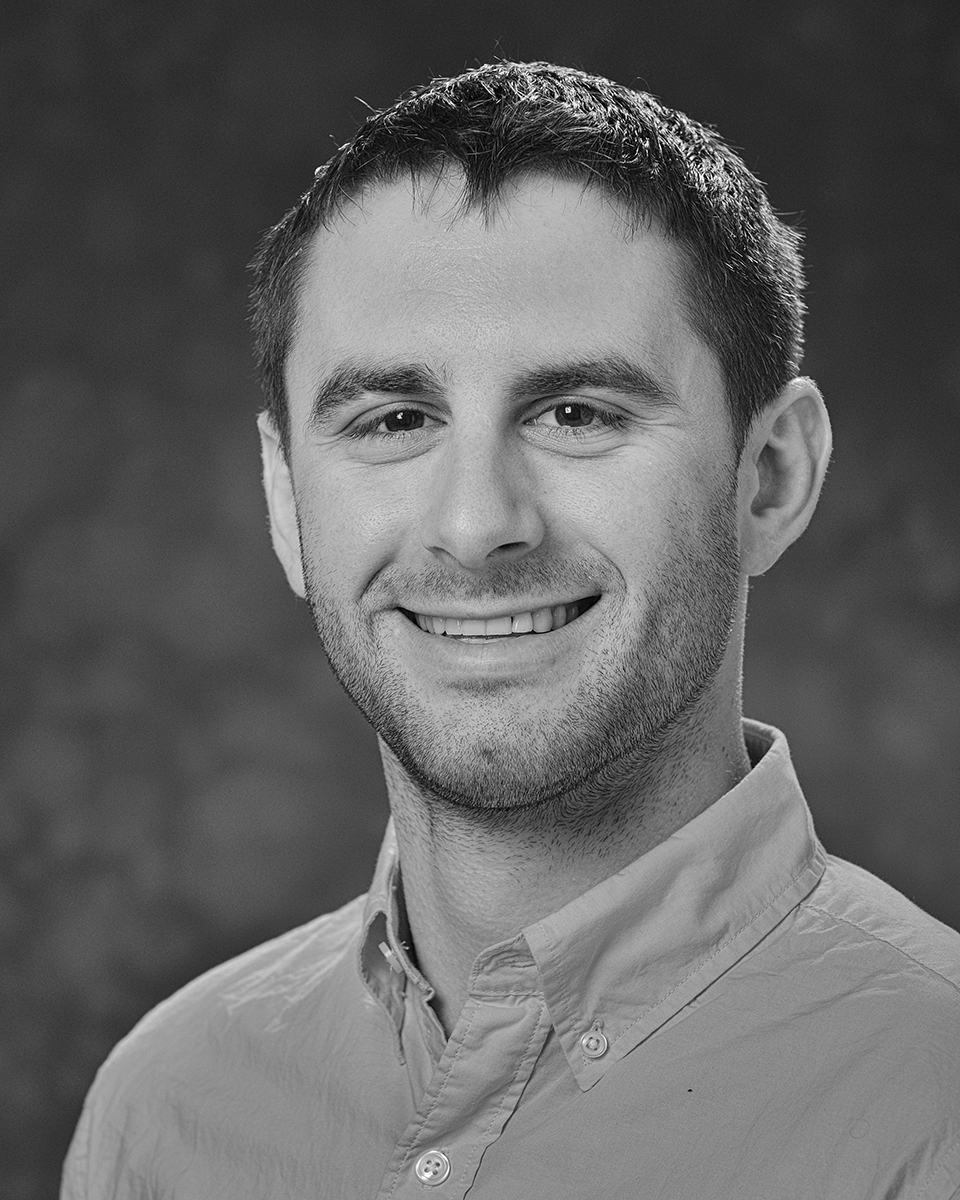}}]{Garrett Bernstein}
is a Computer Science Ph.D. student at the University of
Massachusetts Amherst. At MIT Lincoln Laboratory, he was a member of
the technical staff in the Intelligence and Decision Technologies
Group. His research focused on statistical inference and machine
learning applied to diverse problems, such as graph detection
algorithms, model simulation, semantic analysis, and military
operational effectiveness. Prior to joining the Laboratory, he
received a bachelor's degree in applied and engineering physics and
and engineering master's degree in computer science, both from Cornell
University.
\end{IEEEbiography}

\begin{IEEEbiography}[{\includegraphics[width=1in,clip,keepaspectratio]{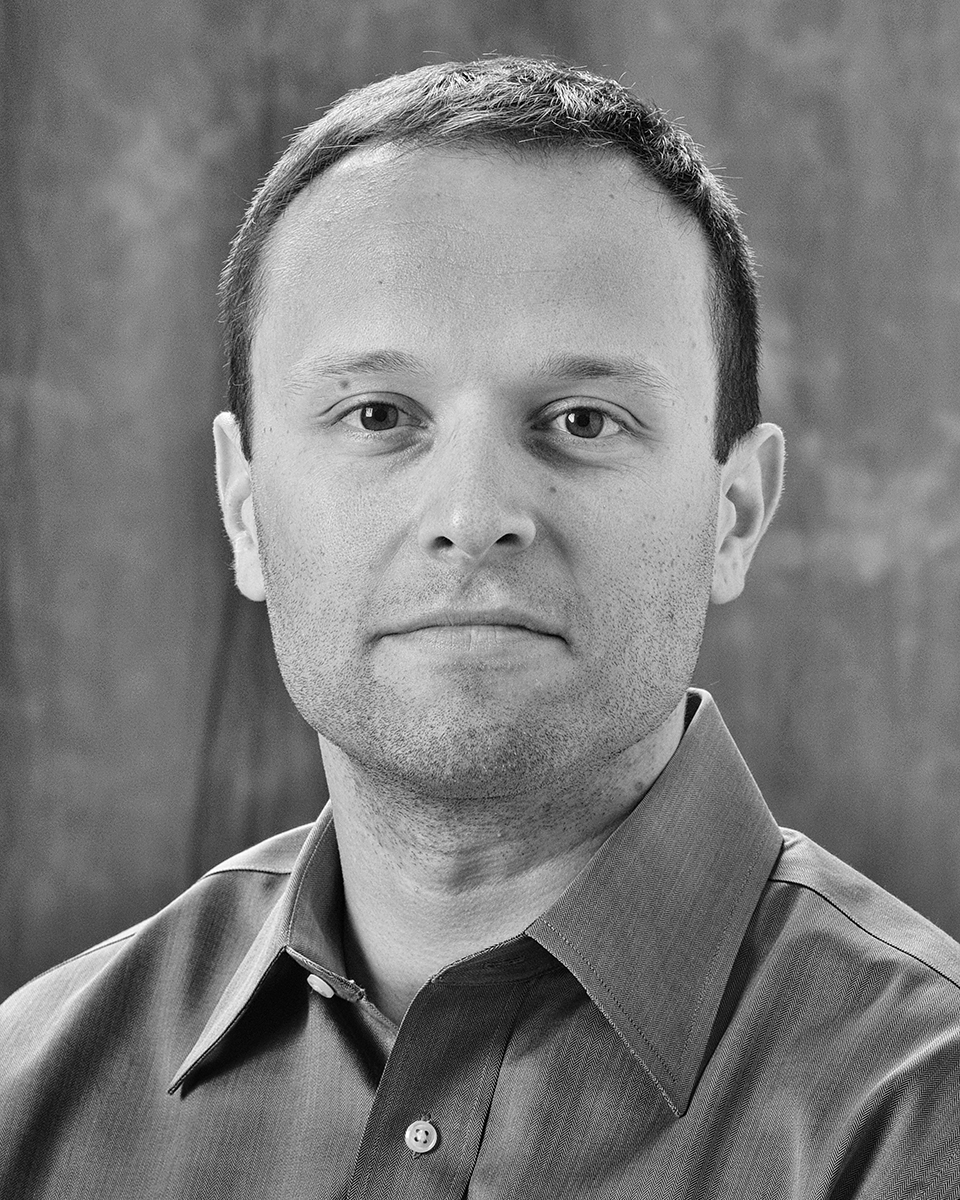}}]{Scott Philips}
is currently a data scientist at Palantir Technologies. Before joining
Palantir, Scott spent five years as a member of the technical staff in
the Intelligence and Decision Technologies Group.  While he was at
Lincoln Laboratory, his research focused on developing statistical
algorithms for the exploitation of data from intelligence,
surveillance, and reconnaissance sensors. Scott received his doctoral
degree in electrical engineering in 2007 from the University of
Washington, where his research focused on signal processing and
machine learning algorithms for the detection and classification of
sonar signals.
\end{IEEEbiography}

\vfill

\end{document}